\documentstyle[12pt]{article}

\advance \topmargin by -\headheight
\advance \topmargin by -\headsep

\evensidemargin \oddsidemargin
\begin{document}

\def\rf#1{(\ref{eq:#1})}
\def\lab#1{\label{eq:#1}}
\def\nonu{\nonumber}
\def\br{\begin{eqnarray}}
\def\er{\end{eqnarray}}
\def\be{\begin{equation}}
\def\ee{\end{equation}}
\def\eq{\!\!\!\! &=& \!\!\!\! }
\def\foot#1{\footnotemark\footnotetext{#1}}
\def\lb{\lbrack}
\def\rb{\rbrack}
\def\llangle{\left\langle}
\def\rrangle{\right\rangle}
\def\blangle{\Bigl\langle}
\def\brangle{\Bigr\rangle}
\def\llb{\left\lbrack}
\def\rrb{\right\rbrack}
\def\lcurl{\left\{}
\def\rcurl{\right\}}
\def\({\left(}
\def\){\right)}
\newcommand{\nit}{\noindent}
\newcommand{\ct}[1]{\cite{#1}}
\newcommand{\bi}[1]{\bibitem{#1}}
\def\lskip{\vskip\baselineskip\vskip-\parskip\noindent}
\relax
\def\mskp{\par\vskip 0.3cm \par\noindent}
\def\sskp{\par\vskip 0.15cm \par\noindent}
\def\tr{\mathop{\rm tr}}
\def\Tr{\mathop{\rm Tr}}
\def\v{\vert}
\def\bv{\bigm\vert}
\def\Bgv{\;\Bigg\vert}
\def\bgv{\bigg\vert}
\newcommand\partder[2]{{{\partial {#1}}\over{\partial {#2}}}}
\newcommand\funcder[2]{{{\delta {#1}}\over{\delta {#2}}}}
\newcommand\Bil[2]{\Bigl\langle {#1} \Bigg\vert {#2} \Bigr\rangle}  
\newcommand\bil[2]{\left\langle {#1} \bigg\vert {#2} \right\rangle} 
\newcommand\me[2]{\left\langle {#1}\right|\left. {#2} \right\rangle} 
\newcommand\sbr[2]{\left\lbrack\,{#1}\, ,\,{#2}\,\right\rbrack}
\newcommand\pbr[2]{\{\,{#1}\, ,\,{#2}\,\}}
\newcommand\pbbr[2]{\lcurl\,{#1}\, ,\,{#2}\,\rcurl}
\def\a{\alpha}
\def\b{\beta}
\def\d{\delta}
\def\D{\Delta}
\def\eps{\epsilon}
\def\vareps{\varepsilon}
\def\g{\gamma}
\def\G{\Gamma}
\def\grad{\nabla}
\def\h{{1\over 2}}
\def\l{\lambda}
\def\L{\Lambda}
\def\m{\mu}
\def\n{\nu}
\def\o{\over}
\def\om{\omega}
\def\O{\Omega}
\def\p{\phi}
\def\P{\Phi}
\def\pa{\partial}
\def\pr{\prime}
\def\ra{\rightarrow}
\def\s{\sigma}
\def\S{\Sigma}
\def\t{\tau}
\def\th{\theta}
\def\Th{\Theta}
\def\ti{\tilde}
\def\wti{\widetilde}
\def\jc{J^C}
\def\bj{{\bar J}}
\def\sj{{\jmath}{}}
\def\bsj{{\bar \jmath}{}}
\def\bp{{\bar \p}}
\def\faa{Fa\'a di Bruno~}
\def\ca{{\cal A}}
\def\cb{{\cal B}}
\def\ce{{\cal E}}
\def\cd{{\cal D}}
\newcommand\sumi[1]{\sum_{#1}^{\infty}}   
\newcommand\fourmat[4]{\left(\begin{array}{cc}  
{#1} & {#2} \\ {#3} & {#4} \end{array} \right)}
\newcommand\twocol[2]{\left(\begin{array}{c}  
{#1} \\ {#2} \end{array} \right)}
\def\lie{{\cal G}}
\def\dlie{{\cal G}^{\ast}}
\def\elie{{\widetilde \lie}}
\def\edlie{{\elie}^{\ast}}
\def\hlie{{\cal H}}
\def\wlie{{\widetilde \lie}}
\def\f#1#2#3 {f^{#1#2}_{#3}}
\def\winf{{\sf w_\infty}}
\def\win1{{\sf w_{1+\infty}}}
\def\hwinf{{\sf {\hat w}_{\infty}}}
\def\Winf{{\sf W_\infty}}
\def\Win1{{\sf W_{1+\infty}}}
\def\hWinf{{\sf {\hat W}_{\infty}}}
\def\cB{{\cal B}}
\def\cH{{\cal H}}
\def\cL{{\cal L}}
\def\cA{{\cal A}}
\def\rlx{\relax\leavevmode}
\def\inbar{\vrule height1.5ex width.4pt depth0pt}
\def\IZ{\rlx\hbox{\sf Z\kern-.4em Z}}
\def\IR{\rlx\hbox{\rm I\kern-.18em R}}
\def\IC{\rlx\hbox{\,$\inbar\kern-.3em{\rm C}$}}
\def\one{\hbox{{1}\kern-.25em\hbox{l}}}
\def\0#1{\relax\ifmmode\mathaccent"7017{#1}%
        \else\accent23#1\relax\fi}
\def\omz{\0 \omega}
\def\ltimes{\mathrel{\vrule height1ex}\joinrel\mathrel\times}
\def\rtimes{\mathrel\times\joinrel\mathrel{\vrule height1ex}}
\def\mark{\noindent{\bf Remark.}\quad}
\def\prop{\noindent{\bf Proposition.}\quad}
\def\theor{\noindent{\bf Theorem.}\quad}
\def\name{\noindent{\bf Definition.}\quad}
\def\exam{\noindent{\bf Example.}\quad}
\def\proof{\noindent{\bf Proof.}\quad}
\def\lemma{\noindent{\bf Lemma.}\quad}
\newcommand\PRL[3]{{\sl Phys. Rev. Lett.} {\bf#1} (#2) #3}
\newcommand\NPB[3]{{\sl Nucl. Phys.} {\bf B#1} (#2) #3}
\newcommand\NPBFS[4]{{\sl Nucl. Phys.} {\bf B#2} [FS#1] (#3) #4}
\newcommand\CMP[3]{{\sl Commun. Math. Phys.} {\bf #1} (#2) #3}
\newcommand\PRD[3]{{\sl Phys. Rev.} {\bf D#1} (#2) #3}
\newcommand\PLA[3]{{\sl Phys. Lett.} {\bf #1A} (#2) #3}
\newcommand\PLB[3]{{\sl Phys. Lett.} {\bf #1B} (#2) #3}
\newcommand\JMP[3]{{\sl J. Math. Phys.} {\bf #1} (#2) #3}
\newcommand\PTP[3]{{\sl Prog. Theor. Phys.} {\bf #1} (#2) #3}
\newcommand\SPTP[3]{{\sl Suppl. Prog. Theor. Phys.} {\bf #1} (#2) #3}
\newcommand\AoP[3]{{\sl Ann. of Phys.} {\bf #1} (#2) #3}
\newcommand\PNAS[3]{{\sl Proc. Natl. Acad. Sci. USA} {\bf #1} (#2) #3}
\newcommand\RMP[3]{{\sl Rev. Mod. Phys.} {\bf #1} (#2) #3}
\newcommand\PR[3]{{\sl Phys. Reports} {\bf #1} (#2) #3}
\newcommand\AoM[3]{{\sl Ann. of Math.} {\bf #1} (#2) #3}
\newcommand\UMN[3]{{\sl Usp. Mat. Nauk} {\bf #1} (#2) #3}
\newcommand\FAP[3]{{\sl Funkt. Anal. Prilozheniya} {\bf #1} (#2) #3}
\newcommand\FAaIA[3]{{\sl Functional Analysis and Its Application} {\bf #1}
(#2) #3}
\newcommand\BAMS[3]{{\sl Bull. Am. Math. Soc.} {\bf #1} (#2) #3}
\newcommand\TAMS[3]{{\sl Trans. Am. Math. Soc.} {\bf #1} (#2) #3}
\newcommand\InvM[3]{{\sl Invent. Math.} {\bf #1} (#2) #3}
\newcommand\LMP[3]{{\sl Letters in Math. Phys.} {\bf #1} (#2) #3}
\newcommand\IJMPA[3]{{\sl Int. J. Mod. Phys.} {\bf A#1} (#2) #3}
\newcommand\AdM[3]{{\sl Advances in Math.} {\bf #1} (#2) #3}
\newcommand\RMaP[3]{{\sl Reports on Math. Phys.} {\bf #1} (#2) #3}
\newcommand\IJM[3]{{\sl Ill. J. Math.} {\bf #1} (#2) #3}
\newcommand\APP[3]{{\sl Acta Phys. Polon.} {\bf #1} (#2) #3}
\newcommand\TMP[3]{{\sl Theor. Mat. Phys.} {\bf #1} (#2) #3}
\newcommand\JPA[3]{{\sl J. Physics} {\bf A#1} (#2) #3}
\newcommand\JSM[3]{{\sl J. Soviet Math.} {\bf #1} (#2) #3}
\newcommand\MPLA[3]{{\sl Mod. Phys. Lett.} {\bf A#1} (#2) #3}
\newcommand\JETP[3]{{\sl Sov. Phys. JETP} {\bf #1} (#2) #3}
\newcommand\JETPL[3]{{\sl  Sov. Phys. JETP Lett.} {\bf #1} (#2) #3}
\newcommand\PHSA[3]{{\sl Physica} {\bf A#1} (#2) #3}
\newcommand\PHSD[3]{{\sl Physica} {\bf D#1} (#2) #3}
\newcommand{\sect}[1]{\setcounter{equation}{0}\section{#1}}
\renewcommand{\theequation}{\thesection.\arabic{equation}}
\relax
\def\alie{{\hat {\cal G}}}
\def\cK{{\cal K}}
\def\cM{{\cal M}}
\def\cR{{\cal R}}
\def\cKP{{\sf cKP}~}
\def\GNLS{{\sf GNLS}}
\def\tv{\ti v}
\def\tu{\ti u}
\newcommand\Back{B\"{a}cklund}
\def\phanta{\phantom{aaaaaaaaaaaaaaa}}

\begin{titlepage}
\vspace*{-1cm}
\noindent
\phantom{bla}
\hfill{UICHEP-TH/95-1}\\
\phantom{bla}
\hfill{hep-th/9503211}
\\
\begin{center}
{\large {\bf Integrable Lax Hierarchies, their Symmetry Reductions \\
and Multi-Matrix Models
\footnotemark
\footnotetext{Lectures presented at the VIII J.A. Swieca Summer School,
Section: Particles and Fields, Rio de Janeiro - Brasil - February/95}}}
\end{center}
\vskip .3in
\begin{center}
{ H. Aratyn\footnotemark
\footnotetext{Work supported in part by the U.S. Department of Energy
under contract DE-FG02-84ER40173}}
\par \vskip .1in \noindent
Department of Physics \\
University of Illinois at Chicago\\
845 W. Taylor St.\\
Chicago, IL 60607-7059\\
{\em e-mail}: aratyn@uic.edu
\end{center}

\vskip .7in

\begin{abstract} \noindent
\noindent
Some new developments in constrained Lax integrable systems and their
applications to physics are reviewed.
After summarizing the tau function construction of the KP hierarchy
and the basic concepts of the symmetry of nonlinear equations,
more recent ideas dealing with constrained KP models are described.
A unifying approach to constrained KP hierarchy based on graded
$SL(r+n,n)$ algebra is presented and equivalence formulas are obtained
for various pseudo-differential Lax operators appearing in this context.

It is then shown how the Toda lattice structure emerges from
constrained KP models via canonical Darboux-B\"{a}cklund transformations.
These transformations enable us to find simple Wronskian solutions for
the underlying tau-functions.

We also establish a relation between two-matrix models and constrained Toda
lattice systems and derive from this relation
expressions for the corresponding partition function.
\end{abstract}

\end{titlepage}

\tableofcontents

\sect{Introduction}
\subsection{Abbreviations; Diagram of Lectures}

{\bf Abbrevations used in text:}

\mskp
KP hierarchy (equation)= Kadomtsev-Petviashvili hierarchy (equation)\\
\cKP hierachy = constrained KP hierarchy\\
DB transfromation = Darboux-B\"{a}cklund transfromation\\
KdV hierarchy (equation)= Korteveg-de Vries hierarchy (equation)\\
BA function= Baker-Akhiezer function\\
AKNS hierarchy = Ablowitz-Kaup-Newell-Segur hierarchy\\
AKS approach = Adler-Kostant-Symes approach\\
GNLS equation= generalized non-linear Schr\"{o}dinger equation

\unitlength=1.00mm
\special{em:linewidth 0.4pt}
\linethickness{0.4pt}
\begin{picture}(130.00,145.00)(0,50)
\put(10.00,130.00){\framebox(40.00,15.00)[cc]{KP Hierarchy}}
\put(90.00,130.00){\framebox(40.00,15.00)[cc]{2-Matrix Model}}
\put(30.00,130.00){\vector(0,-1){30.00}}
\put(110.00,130.00){\vector(0,-1){30.00}}
\put(10.00,85.00){\framebox(40.00,14.00)[cc]{$sl(r+n,n)$ \cKP}}
\put(10.00,70.00){\framebox(40.00,15.00)[cc]{$sl(r)$ KdV}}
\put(90.00,85.00){\framebox(40.00,14.00)[cc]{constr. Toda Lattice}}
\put(51.00,93.00){\vector(1,0){39.00}}
\put(72.00,95.00){\makebox(0,0)[cc]{DB}}
\put(59.00,91.00){\makebox(0,0)[lc]{transformation}}
\put(10.00,115.00){\makebox(0,0)[lc]{symmetry}}
\put(32.00,115.00){\makebox(0,0)[lc]{constraint}}
\put(107.00,115.00){\makebox(0,0)[rc]{string}}
\put(114.00,115.00){\makebox(0,0)[lc]{equation}}
\put(50.00,93.00){\vector(1,0){40.00}}
\put(90.00,93.00){\vector(-1,0){40.00}}
\put(50.00,93.00){\vector(1,0){40.00}}
\put(90.00,93.00){\vector(-1,0){40.00}}
\put(50.00,93.00){\vector(1,0){40.00}}
\put(90.00,93.00){\vector(-1,0){40.00}}
\put(70.00,165.00){\makebox(0,0)[ct]{{\large Fig. 1.}  {\large
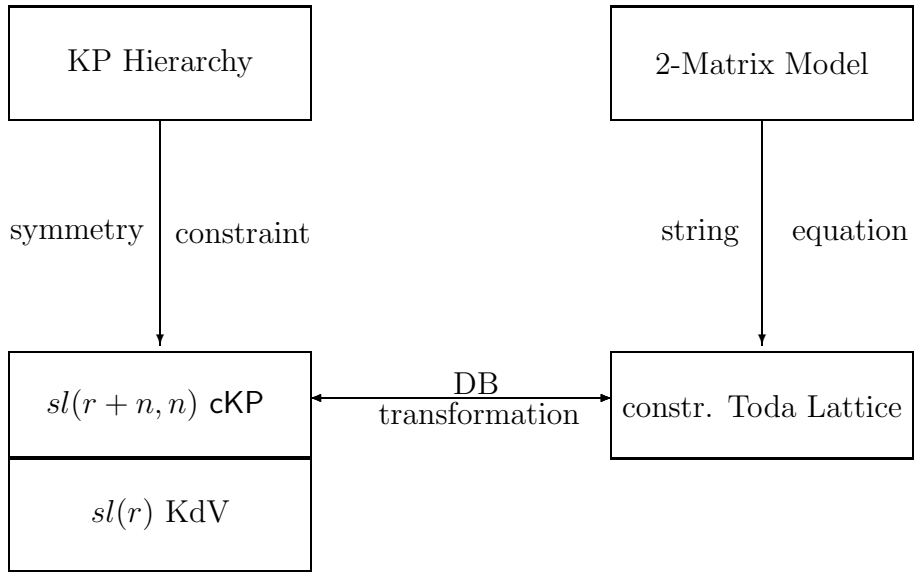
{\bf Structure of exposition}}}}
\end{picture}

\subsection{Content of Lectures}

In these lectures we study constrained KP hierarchy, its connection to
the discrete Toda models and application via Toda models to the multi-matrix
approach to the two-dimensional gravity.

A presentation follows the diagram shown on the previous page in the
counter-clockwise direction.
In Section 2 we collect some standard facts on KP hierarchy which are used in
several places in these lectures.
A link between the BA function and $\t$ function  is established and
it is shown how to obtain the conserved current densities
of the KP hierarchy from these concepts.
A complementary presentation of KP hierachy centered around
multi-Hamiltonian/Poisson structures was given at the previous Swieca school
\ct{swieca-z}.

The notion of constrained KP hierarchies is introduced in Section 3.
Since this notion is significant for our lectures we introduce
a variety of reduction schemes and elaborate on their mutual equivalency.

The picture which emerges from this discussion is most simply presented in
the language of pseudo-differential operators.
Two basic equivalent approaches look especially attractive and
deserve to be mentioned here in the introduction.

One construction introduces the Lax operator of the constrained KP (\cKP)
hierarchy as a ratio
\be
L_{m,n} \equiv { L^{(m)} \o  L^{(n)} }  \qquad \; n\leq m-1
\lab{i-ratio-a}
\ee
of two purely differential operators
\be
L^{(m)} = \( D + v_m\) \( D + v_{m-1} \) \cdots \( D + v_1 \)
\quad ;\quad
L^{(n)} = \( D + \tv_{n} \) \( D + \tv_{n-1} \) \cdots \( D + \tv_1 \)
\lab{i-lmk-a}
\ee
with coefficients $v_i, \tv_i$ subject to a constraint:
\be
\sum_{j=1}^{m} v_j - \sum_{l=1}^{n}  \tv_l= 0
\lab{i-psimk1}
\ee
Imposing condition \rf{i-psimk1} is equivalent
to requiring tracelessness of the underlying graded
$SL (m,n)$ algebra.
For this reason we refer to this class of constrained KP (\cKP)
models as $SL (m,n)$ \cKP hierarchy.
This construction is useful to describe a bi-Poisson structure of
the \cKP hierarchy and in fact involves variables which almost
abelianize the second Poisson structure.
There is another formulation of the same class of models which is based
on an alternative expression for the Lax operator given in \rf{i-ratio-a}:
\be
L_{m=r+n,n} = D^r+ \sum_{l=0}^{r-2} u_l D^l +
\sum_{i=1}^n \Phi_i D^{-1} \Psi_i
\lab{i-ratio-b}
\ee
A connection between variables from \rf{i-ratio-a} and \rf{i-ratio-b}
takes a form of complicated Miura-like transformations ( see Section 3 and
\ct{office}), but for $\Psi_i$'s the link is very simple since
$\Psi_i$'s turn out to be elements of the $n$-dimensional kernel of $L^{(n)}$.
Furthermore $\Phi_i , \Psi_i$ abelianize the first bracket structure.
Their property of being eigenfunctions of $L_{r+n,n}$ makes the expression
in \rf{i-ratio-b} a proper framework for discussing the Lax equation in
the setting of the $SL (m,n)$ \cKP hierarchy.

To go further and find $\t$-functions corresponding to the
$SL (m,n)$ \cKP hierarchy we need a notion of the Darboux-B\"{a}cklund
(DB) transformations which we introduce in Section 4.
With this notion we are able to derive simple Wronskian expressions for
the $\t$-functions.
We also reveal a connection between on one side (constrained) Toda discrete
models and on another \cKP models endowed with DB symmetry structures.
The picture which emerges is of the continuous integrable model with canonical
symmetry mimicking the lattice shift of the corresponding Toda lattice.
This scenario has been appearing in various forms
in literature \ct{chud,shabat-91,
lez-can,BX9212,BK88,fla83a,ne85,discrete,similar}.

The observations of Section 4 provide a right scene for establishing
the link to two-matrix model.
The two-matrix model is rewritten in Section 5 as a linear discrete system
with additional constraint; a string equation.
Using the string equation we are able to cast the lattice equations
into the form of a constrained Toda equation which can be represented
via construction of Section 4 to the class of \cKP models.

We have included two appendices on Schur polynomials and Wronskians,
which may clarify few technical points.
\subsection{Acknowledgements }
These lectures have grown out of work I have done over the past few years
with L.A. Ferreira, J.F. Gomes,
E. Nissimov, S. Pacheva and A.H. Zimerman.
I thank them for sharing their insights with me.

It is a pleasure to thank organizers of the School for their warm
hospitality and giving me opportunity to present these lectures.
\newpage
\sect{Pseudo-differential Operators, BA Functions and Hirota Equations}
\subsection{Lax Equation and Eigenfunction for the Lax Operator}
We begin with the Sato theory of the KP
hierarchy.
Let $ \{ t_j \} $ denote a set of independent variables
with $ t_1\equiv x $.
The formulation of the KP hierarchy  is based on the Lax equations
\be
\partder{L}{t_n} = \lb B_n \, , \, L \rb
\quad    \; \; n = 1, 2, \ldots \lab{lax-eq}
\ee
describing isospectral deformations of the pseudo-differential operator:
\be
L = \pa+ \sumi{i=0} u_i \pa^{-i-1} \lab{lax-op}
\ee
Here $ u_n $ are the functions of $ \{ t_j \} $, and
$ B_n = ( L^n )_{+} $ is the truncation to  the differential part of $ L^n $.
A negative power part of the differential operator $ L^n $ will be
written as $ ( L^n )_- $.
The Lax equations \rf{lax-eq} have the zero-curvature representation
taking the Zakharov-Shabat form:
\be
  \frac{\partial B_n}{\partial t_m} - \frac{\partial B_m}{\partial t_n}
    + \lb B_n , B_m \rb = 0 ,\; \; \qquad n, m = 1, 2, \ldots
\lab{zs-eq}
\ee

It is often convenient to view \rf{lax-op} and \rf{zs-eq}
as integrability conditions of the linear system:
\br
L \psi (t,\l ) &= &\l \psi(t, \l)  \lab{ba-a} \\
\frac{\pa \psi(t,\lambda)}{\pa t_n} &= &B_n \psi(t, \lambda) \lab{ba-b}
\er
In connection with the above linear eigenvalue problem one introduces two
class of functions.

\name {\em A function $\Phi$ is called \underbar{eigenfunction} for the Lax
 operator $L$ satisfying
Sato's flow equation \rf{lax-eq} if its  flows are given by
expression:}
\be
\partder{\Phi}{t_m} = \( L^m\)_{+} \Phi
\lab{eigenlax}
\ee
{\em for the infinite many times $t_m$.}

A special role is played by an eigenfunction function, which also
satisfies the spectral equation \rf{ba-a}.

\name
{\em A function $ \psi(t,\lambda) $ satisfying relations \rf{ba-a} and
\rf{ba-b} is called a \underbar{Baker-Akhiezer} function.}

\subsection{Conservation Laws of KP Hierarchy,
and Tau-Function Representation of the Baker-Akhiezer Function}

Note first that \rf{lax-op} can be inverted providing an expansion of $D$ in
powers of Lax operator:
\be
D= L + \sumi{i=1} \s_{i}^{(1)} L^{-i}
\lab{painl}
\ee
Similar inversion relation can also be written for the differential operator
$B_m$
\be
B_m = L^m + \sumi{i=1} \s_{i}^{(m)} L^{-i}
\lab{bminl}
\ee
Expand now $\ln \psi(t,\lambda)$ as
\be
\ln\psi(t,\lambda) = \sum_{n=1}^\infty t_n\lambda^n
+\sum_{i=1}^\infty \psi_{i}\lambda^{-i},
\lab{logpsi}
\ee
Because of the linear equation \rf{ba-a} for $L$, one finds
\be
\chi (\l)\equiv  \pa \psi(t,\lambda)  /\psi(t,\lambda)
    = \l +\sumi{i=1} \s_{i}^{(1)} \l^{-i}
\lab{expsi}
\ee
where we have defined for convenience an auxiliary quantity $\chi=
\pa \ln\psi$.
Hence by comparing \rf{painl} and \rf{logpsi} we get
$ \pa \psi_{i} = \s_{i}^{(1)}$.

The conservation laws for the KP hierarchy in form we will discuss here
follow from the recurrence relation for the differential operators
$B_m$. From \rf{bminl} it follows namely
\br
B_{m+1} \eq \( D - \sumi{i=1} \s_{i}^{(1)} L^{-i}\)
L^m + \sumi{i=1} \s_{i}^{(m+1)} L^{-i}  \nonu \\
\eq \( D - \sumi{i=1} \s_{i}^{(1)} L^{-i}\)
\( B_m - \sumi{j=1} \s_{j}^{(m)} L^{-j} \) + \sumi{i=1} \s_{i}^{(m+1)} L^{-i}
\lab{bmrec}\\
\eq D B_m - \sum_{j=1}^m \s_{j}^{(1)} B_{m-j} - \s_{1}^{(m)}
+ O \,( D^{-1})
\nonu
\er
Since the left hand side is a pure differential operator it is obvious
that terms with $D^{-k}$ with $k > 0$ cancel and consequently the only
non-zero contribution comes from the first three terms on the
right hand side.
The following identity:
\be
\llb D - \chi (\l) \rrb \cB (\l) = \eps (\l)
\lab{bmrec-a}
\ee
with
\be
\eps (\l) \equiv -1 + \sumi{m=1} \l^{-m-1} \s_{1}^{(m)} \qquad ;\qquad
\cB (\l) \equiv \sumi{m=0} \l^{-m-1} B_m \nonu
\ee
presents a compact way of expressing the recurrence relation
of equation \rf{bmrec}.
We now follow reference \ct{fla83b} (see also \ct{wilson81,chered}) and
apply \rf{bmrec-a} on the BA function $\psi (z) $ with expansion parameter
being $z$ instead of $\l$:
\be
\sumi{m=0} \l^{-m-1} \pa B_m \psi (z) - \chi (\l)
\sumi{m=0} \l^{-m-1} B_m \psi (z) = \eps (\l) \psi (z)
\lab{bmrec-b}
\ee
or
\be
\sumi{m=0} \l^{-m-1} \( \pa B_m \psi (z) \)  \cdot \psi^{-1} (z)
- \chi (\l) \sumi{m=0} \l^{-m-1}
 B_m \psi (z) \cdot \psi^{-1} (z)  = \eps (\l) \lab{bmrec-c}
\ee
Using that $B_m \psi (z) \cdot \psi^{-1} (z) = z^m + \s_{1}^{(m)} z^{-1}
+ \ldots$ we obtain
\br
\eps (\l) &=&\sumi{m=0} \l^{-m-1} \pa \(B_m \psi (z) \cdot \psi^{-1} (z) \)
- {\chi (z) -\chi (\l ) \o z- \l}   \lab{bmrec-d}\\
&- & \chi (\l) \sumi{m=0} \l^{-m-1} \(\s_{1}^{(m)} z^{-1} + \ldots\)
+ \chi (z) \sumi{m=0} \l^{-m-1} \(\s_{1}^{(m)} z^{-1}
+ \ldots\)\nonu
\er
which upon taking the limit $z \to \l$ yields
\be
\partder{\chi(\l)}{\l} + \eps (\l) = \pa \sumi{m=0} \l^{-m-1}
\(B_m \psi (\l ) \cdot \psi^{-1} (\l) \)
\lab{derchi}
\ee
Hence modes of the left hand side of \rf{derchi}:
$ \s_{1}^{(l)} -l \s_{l}^{(1)} \; , \; l=1, \ldots$ are
conserved densities of the KP hierarchy.
{}From
\be
 \pa \(B_m \psi \cdot \psi^{-1} \)  =
 \pa \(\partder{\psi}{t_m} \cdot \psi^{-1} \)  =
  \partder{}{t_m}  \(\pa \psi \cdot \psi^{-1} \) =   \partder{\chi}{t_m}
\lab{derchi-a}
\ee
we find that also $\pa \s_{l}^{(1)} / \pa t_m$ define
conserved densities of the KP hierarchy and moreover comparing \rf{derchi}
and \rf{derchi-a} we get a relation between these conserved quantities:
\be
\sum_{j=1}^{m-1} \partder{\s_{m-j}^{(1)}}{t_j} =
\s_{1}^{(m)} - m\, \s_{m}^{(1)}
\lab{conserv}
\ee
{}From an obvious identity
\be
0 = \( { \pa^2 \psi \o \pa t_k \pa t_j } -{ \pa^2 \psi \o \pa t_j \pa t_k } \)
\psi^{-1} = \( { \pa \s_1^{(j)}\o \pa t_k } -  { \pa \s_1^{(k)}\o \pa t_j }
\) \l^{-1} + \ldots
\lab{exact}
\ee
we see that we can write $ \s_1^{(m)} =\pa f/ \pa t_m$ with some arbitrary
function $f$, which we will choose to write as $ f = - \pa_x \ln \tau$.
Hence in this new notation $ \s_1^{(m)} =-\pa_x \pa_m \ln \tau$
and clearly $ \pa_n   \s_1^{(m)} =  \pa_m   \s_1^{(n)} $.
Furthermore we can now rewrite \rf{conserv} as
\be
\partder{}{t_m} \pa_x \ln \tau = - m  \s_{m}^{(1)} - \sum_{k=1}^{m-1}
\partder{\s_{m-k}^{(1)}}{t_k}
\lab{conserv-a}
\ee
\mark
{}From \rf{painl} $\s_1^{(1)} = - {\rm Res}_{\pa} (L)$ and therefore
$ u_0 = \pa_x^2 \ln \tau$.
Recall that for the pseudo-differental
operator $A= a_1 D^{-1} + \ldots$ we have ${\rm Res}_{\pa} A =a_1$.

On basis of identity \rf{schur-le} (or \rf{schur-lea}) for the Schur
polynomials $p_n$ shown in Appendix A one verifies that \rf{conserv-a}
has a solution given by
\be
 \s_{n}^{(1)} = \pa_x p_n ( - {\wti \pa} ) \ln \tau \quad;\quad n \geq 1
\lab{sigmam}
\ee
where
\be
{\ti \pa} \equiv \( \partder{}{t_1}, {\pa \o 2 \pa t_2}, {\pa \o 3\pa t_3} ,
\ldots \)
\lab{tipa}
\ee
Comparing with \rf{logpsi} and recalling that $\pa \psi_i= \s_i^{(1)}$
we see that (up to an integration constant)
\be
\ln \psi(t,\lambda) = \sum_{n=1}^\infty t_n\lambda^n
+\sum_{i=1}^\infty \lambda^{-i} p_i ( - {\wti \pa} ) \ln \tau
\lab{logpsi-a}
\ee
which yields the main result of this subsection:
\br
\psi(t,\lambda) &=& e^{\xi(t,\lambda)}\, { \exp \( - \sumi{l=1}  \l^{-l}
{\pa \o l \pa t_l} \) \tau (t)\o \tau (t)}
 = e^{\xi(t,\lambda)} \sumi{n=0} { p_n \( - {\ti \pa}\)
\tau (t)\o \tau (t)} \l^{-n} \lab{psi-main} \\
&=& e^{\xi(t,\lambda)} { \tau \(t_i - 1/i \l^{i}\)\o \tau (t_i)}  \nonu
\er
where
\be
\xi(t,\lambda) \equiv  \sum_{n=1}^\infty t_n\lambda^n
\lab{xidef}
\ee
\subsection{Dressing Operator}
It is possible to reproduce the Lax operator \rf{lax-op} through the dressing
formula
\be
L = W D W^{-1}
\lab{dress}
\ee
where the dressing operator $ W $ is the pseudo-differential
operator:
\be
W = 1 + \sumi{i=1} w_i (t) D^{-i}
\lab{dress-op}
\ee
satisfying the Sato equations:
\br
\frac{\partial W}{\partial t_n} &=& B_n W - W D^n
= - \( L^n\)_{-} W
\lab{sato-a}\\
L W &=& W D.
\lab{sato-b}
\er
{}From the dressing formula it follows immediately that the BA
function  can be rewritten as
\be
\psi (t,\lambda) = W \exp ( \sum_{n=1}^\infty t_n \lambda^n )
= {\hat w} (t,\lambda) \, e^{\xi (t,\lambda) }
\lab{psi-dress}
\ee
where
\be
{\hat w} (t,\lambda) = { \tau \(t_i - 1/i \l^{i}\) \o \tau \(t_i \)}
\lab{hatw}
\ee
as can be found from \rf{psi-main}.

In a similar way an adjoint BA function $\psi^{*} (t,\lambda)$
can be constructed as follows:
\br
\psi^{*} (t,\lambda) &=& (W^{-1})^{*} e^{-\xi (t,\lambda) }
= {\hat w}^{*} (t,\lambda) e^{-\xi (t,\lambda) }
\lab{psi-star} \\
{\hat w}^{*} (t,\lambda) &=& { \tau \(t_i + 1/i \l^{i}\) \o \tau \(t_i \)}
\lab{hatwsta}
\er
and defines a linear system:
\br
L^{*} \psi^{*} (t,\l ) &= &\l \psi^{*}(t, \l)  \lab{staba-a} \\
\frac{\pa \psi^{*}(t,\lambda)}{\pa t_n} &=& - B_n^{*} \psi^{*}
(t, \lambda) \lab{staba-b}
\er
where $L^{*}$ is adjoint of $L$ and $B_n^{*} $ is a differential part of
$(L^{*})^n$.
Recall that the adjoint operator $A^{\ast}$ of $A= a D^k+ \ldots$ is given by
$A^{\ast}= (-1)^k D^k a + \ldots$.

\subsection{Hirota Equations}
As seen from \rf{psi-dress} and \rf{psi-main} we can rewrite the dressing
operator as
\be
W = \sumi{i=0} {p_i (- {\ti \pa}) \tau (t) \o \tau (t)} \, D^{-i}
\lab{dress-b}
\ee
leading to
\be
L^n = W D^n W^{-1} = \sumi{i,j}  {p_i (- {\ti \pa}) \tau (t) \o \tau (t)}
D^{n-i-j} {p_j ({\ti \pa}) \tau (t) \o \tau (t)}
\lab{ln}
\ee
Let us insert the above expressions into the Sato equation \rf{sato-a}.
Noticing that
\be
-(L^n)_{-} = -\sum_{i,j\geq 0}^{i+j=n+1} {p_i (- {\ti \pa})\tau (t) \cdot
p_j ({\ti \pa}) \tau (t) \o \tau^2 (t) } D^{-1} + \ldots
\lab{lnminus}
\ee
and taking the residue on both sides of \rf{sato-a} we get:
\be
\tau {\pa^2 \tau \o \pa t_n \pa t_1} - \partder{\t}{t_n} \partder{\t}{t_1} -
\sum_{i,j\geq 0}^{i+j=n+1} p_i (- {\ti \pa})\tau (t) \cdot
p_j ({\ti \pa}) \tau (t) =0
\lab{hiro-a}
\ee
or
\be
\(  \h D_1 D_n - p_{n+1} ( {\ti D}) \) \t \cdot \t =0
\lab{hiro-b}
\ee
where we used Hirota's operators defined by
\be
D^m_j a \cdot b = {\pa^m \o \pa s^m_j} a (t_j + s_j) b (t_j - s_j) \v_{s_j=0}
\lab{hiro-ope}
\ee
The first non-trivial eqution obtained from \rf{hiro-b} for $n=3$ ( by
dropping odd polynomials in $D$, which do not make any non-zero contribution)
is the KP equation:
\be
\( D_1^4 + 3 D_2^2 - 4 D_1 D_3 \) \t \cdot \t =0
\lab{hiro-kp}
\ee
which in components takes the usual form:
\be
\partder{}{x} \( \partder{u_0}{t_3}- {1 \o 4} {\pa^3 u_0 \o \pa x^3}
-3 u_0 \partder{u_0}{x} \) - {3 \o 4} {\pa^2 u_0 \o \pa t_2^2} =0
\lab{kp-eq}
\ee
with $u_0 = {\rm Res}_{\pa} L= \pa_x^2 \ln \tau$.

\mark Eq. \rf{hiro-b} is contained in the Hirota differential
equations for the KP hierarchy taking the following expression:
\be
\sumi{j=0} p_j ( -2y) p_{j+1} ( {\ti D}_t ) \exp \lcurl \sumi{l=1} y_l D_{t_l}
\rcurl \t (t) \cdot \t (t) = 0
\lab{hiro-c}
\ee
with $ y = (y_1, y_2, \ldots)$ being an extra multi-variable.

Indeed, it is easy to see that \rf{hiro-b} can be obtained from \rf{hiro-c}
as a coefficient of the term linear in $y_n$.
It can be shown that the Hirota differential equations \rf{hiro-c}
can be obtained from the bilinear identity for the $\t$ functions:
\be
\int \t \( t_1 -{ 1 \o \l}, t_2 -{ 1 \o 2 \l^2},\ldots \)\;
\t \( t_1^{\pr} + { 1 \o \l}, t_2^{\pr} +{ 1 \o 2 \l^2},\ldots \)
e^{\xi ( t-t^{\pr}, \l) } d \l =0
\lab{bili}
\ee
by expanding arguments of both $\t$-functions in \rf{bili} \ct{date,kac}.
Hence \rf{bili} carries a complete information about the KP hierarchy.
The KP formalism based on \rf{bili}  instead on the pseudo-differential Lax
operators can be analyzed in terms of the free fermions Fock space.
These beautiful results are beyond the current discussion (see for details
\ct{date,kac}).
\subsection{Hamiltonian Densities and the ${\bf \t}$-Function}
Let us now consider the product $s \equiv \psi\psi^{*} $ of the BA function
and its adjoint.
We will now show that $s_x $ generates infinitely many commuting
KP flows.
First as in \ct{matsu} note from \rf{hatw} and \rf{hatwsta} that
\br
\psi\psi^{*} \eq { \tau \(t_i - 1/i \l^{i}\)
\tau \(t_i + 1/i \l^{i}\)
\o \tau^2 \(t_i \)} =
{1 \o \tau^2 \(t_i \)}
\exp \( \sumi{k=1}{ \pa \o k \l^k \pa {\eps_k} }\) \t \(t + \eps \)
\t \(t - \eps  \) \Bgv_{\eps=0}      \nonu \\
\eq {1 \o \t^2 \(t \)} \sumi{k=0} { p_k \( {\ti D} \) \t \cdot \t \o \l^k}
= {1 \o 2 \t^2 \(t \)} \sumi{k=0} { D_1 D_{k-1}\t \cdot \t \o \l^k}
\lab{psipsi}
\er
where in the last equation we have used the Hirota equation \rf{hiro-b}.
This equation suggests to consider the Laurent coefficients of $s$ defined
through
\be
s = \psi\psi^{*}= \sumi{n=0} s_n \l^{-n} \quad \to \quad
s_n = { D_1 D_{n-1}\t \cdot \t \o 2 \t^2 }
\lab{laurents}
\ee
We now recall the following technical lemma \ct{dickey-b}:

\lemma {\em Let $P$, $Q$ be pseudo-differential operators. We have}
\be
{\rm Res}_{\l} \llb \( P e^{x \l} \) \( Q^{\ast} e^{-x \l} \) \rrb
= {\rm Res}_{\pa} \( P Q \)
\lab{tlemma}
\ee
{\em as follows from a direct verification.}

Taking $P=WD^n$ and $Q= W^{-1}$ we obtain from the Lemma that
\be
s_{n+1} = {\rm Res}_{\l} \llb \l^n \psi \psi^{\ast} \rrb = {\rm Res}_{\pa}
\(L^n \)
\lab{snln}
\ee
{}From  \rf{psipsi} we find
\be
s_n = {\rm Res}_{\pa} \(L^{n-1} \) = {1 \o 2 \t^2 \(t \)} D_1 D_{n-1}
\t \cdot \t = \pa_x \pa_{n-1} \ln \t
\lab{sntau}
\ee
Since $ u_0 = \pa^2 \ln \t$ we can rewrite $\pa_x s_n $ as
\be
\pa_x s_n = \partder{u_0}{t_{n-1}}
\lab{pasu}
\ee
Note that in notation of the previous subsections the above equation
shows that the conserved densities of the KP hierarchy $\s^{(m)}_1$ are
equal to Hamiltonian densities $\cH_m \equiv {\rm Res}_{\pa} \(L^{m}\)$ of
the KP hierarchy, which generate commuting flows through relation
$ \pa_x \cH_n = (\pa /\pa t_n) u_0$.
Accordingly the KP hierarchy has an infinite set of commuting symmetries
associated with infinite set of the conserved Hamiltonians.
This translates into commutativity of Hamiltonians on the level of the
Poisson brackets indicating integrability of the system.
These symmetries are called {\em isospectral} symmetries as they preserve
the spectrum of the underlying linear problem.
\newpage
\sect{Symmetry Reduction of the Integrable Systems. Five Constructions of
the \cKP models.}

\subsection{Symmetry of Nonlinear Evolution Equation}
We will discuss notion of symmetry of partial differential equatiom on the
basis of an evolution equation:
\be
\partder{u}{t} = K \lb u (t) \rb
\lab{flow}
\ee
with some nonlinear operator $K$.
An evolution equation $\partder{u}{\s} = G \lb u (t) \rb$
is called a {\sl symmetry} of \rf{flow} if the Fr\'{e}chet derivative of $K$
at the point $u$ in the direction of $G$
\br
K^{\pr} \lb u (t) \rb \( G\) &\equiv&
\partder{}{\eps} K \lb x, u (t)+ \eps G, u^{\pr} +\eps G_x, \ldots
\rb\v_{\eps=0}  \lab{frechet} \\
G_x &\equiv& \partder{G}{x} + \sum_{k \leq 1} \partder{G}{u^{(k)}} u^{(k+1)}
\lab{totder}
\er
satisfies
\be
K^{\pr} \lb u (t) \rb \( G\) \equiv
\partder{G}{t} + G^{\pr} \lb u (t) \rb \( K\)
\lab{symdef}
\ee
Hence a condition that $G$ is a symmetry of \rf{flow} is equivalent to
having $v\equiv G \lb u (t) \rb$ satisfy the {\sl linearized} version
of \rf{flow}  with a background $u (t)$ i.e. $ d v /d t =
K^{\pr} \lb u (t) \rb \( v\) $.
In other words $u+\eps v$ satisfies \rf{flow} for all solutions $u$  of
\rf{flow} for an arbitrarily small $\eps$.
Therefore the linearized version of the nonlinear evolution equation
contains all the pertinent information about its symmetries.

We can also rephrase condition \rf{symdef} as commutativity
$\sbr{D_t}{D_{\s}}=0$ of two differentiations associated with $t$ and $\s$
\br
D_t &\equiv& \partder{}{t} + K  \partder{}{u} + K_x  \partder{}{u_x} + \ldots
\lab{difft} \\
D_{\s} &\equiv& \partder{}{\s} + G  \partder{}{u} + G_x  \partder{}{u_x} +
\ldots \lab{diffsi}
\er
\underbar{Examples:}

Consider the KdV equation: $ u_t = 6 u u^{\pr} -u^{\pr\pr\pr} $
in terms of $\omega^{\pr} = u$:
\be
\omega_t = K \lb \omega \rb= 3 \omega^{\pr \, 2} -  \omega^{\pr\pr\pr}
\lab{kdveq}
\ee
Condition \rf{symdef} gives
\be
d v /d t = K^{\pr} \lb \omega (t) \rb \( v\) =
6 \omega^{\pr } v^{\pr } -  v^{\pr\pr\pr}
\lab{kdveqlin}
\ee
Consider Galilean symmetry of KdV:
\be
\omega \to {\ti \omega} = \omega ( x+ 3 t \eps ) + \h x \eps
\lab{galil}
\ee
which can be cast in the form of ``evolution" equation:
\be
v = G = { d {\ti \omega} \o d \eps} \v_{\eps=0} = 3 t {\omega}^{\pr} + \h x
\lab{galili}
\ee
Taking into account that $v_{x} = 3 t {\omega}^{\pr\pr} + \h $,
$v_{xxx} = 3 t {\omega}^{IV}$ and also
$ d v / d t = 3 {\omega}^{\pr} +3 t \( 3 {\omega}^{\pr\, 2} -
{\omega}^{\pr\pr\pr} \)^{\pr}$ we can verify that the Galilean transformation
is a ``non-isospectral" symmetry of the KdV equation.

For the KP eq. \rf{kp-eq}  its linearized version takes the
form:
\be
d v /d t ={1 \o 4} {\pa^3 v\o \pa x^3}
+ 3 \partder{(u_0 v)}{x}  + {3 \o 4} \pa^{-1} {\pa^2 v \o \pa t_2^2}
\lab{kp-eqlin}
\ee

\subsection{Symmetry Reduction and the \cKP Hierarchy}

We will first show that the quantity $\pa_x (\Phi \Psi)$ is a symmetry of
the KP equation with $\Phi$ and $\Psi $ are (adjoint ) eigenfunctions of
$L$ {\sl i.e.} they satisfy \rf{ba-b} and \rf{staba-b}:
\be
\partder{}{t_m} \Phi= L^{{m}}_{+} \Phi
\quad ; \quad
\partder{}{t_m} \Psi = - {L^{m}}^{\ast}_{+} \Psi
\lab{eigenfcts}
\ee
where ${L^{m}}^{\ast}$ is the adjoint operator of ${L^{m}}$.

Let us particularly stress that the above eigenfunctions do not
need to be Baker-Akhiezer eigenfunctions of $L$ and
$\pa_x (\Phi \Psi)$ in general differs from quantity $s_x \equiv
\pa_x \psi\psi^{*} $, which, as we have seen in the previous section,
generates  an infinite set of commuting symmetries
associated with infinite set of the conserved Hamiltonians of the KP hierarchy.

For $n=2$ equation \rf{eigenfcts} gives:
\be
\partder{\Phi}{t_2} = {\pa^2 \Phi \o \pa x^2} + 2 u_0 \Phi \quad;\quad
\partder{\Psi}{t_2} = -{\pa^2 \Psi\o \pa x^2} - 2 u_0 \Psi
\lab{eigentwo}
\ee
while for $n=3$ we obtain from \rf{eigenfcts}:
\br
\partder{\Phi}{t_3} \eq {\pa^3 \Phi \o \pa x^3} + 3u_0 \partder{\Phi}{x} +
{3 \o 2} u^{\pr}_0 \Phi+ {3 \o 2} \( \pa^{-1}_x \partder{u_0}{t_2} \) \Phi
\nonu \\
\partder{\Psi}{t_3} \eq {\pa^3 \Psi \o \pa x^3} + 3u_0 \partder{\Psi}{x}+
{3 \o 2} u^{\pr}_0 \Psi - {3 \o 2} \( \pa^{-1}_x \partder{u_0}{t_2} \) \Psi
\lab{eigenthree}
\er
Using equations \rf{eigentwo} and \rf{eigenthree} one can show as in
\ct{matsu,carroll93} that $S \equiv \Phi\Psi$ satisfies the evolution equation
\be
\partder{S}{t_3} = {1 \o 4} {\pa^3 S\o \pa x^3} + 3 u_0 \partder{S }{x}
+ {3 \o 4}\pa^{-1} {\pa^2 S_x \o \pa t_2^2}
\lab{lkp-eq}
\ee
and therefore the quantity $S_x \equiv \pa_x S$ satisfies
a linearized version of the KP equation \rf{kp-eqlin}.
Hence according to the above definition $S_x$ is a symmetry of the
KP equation.

Recall now that in components the Lax equation \rf{lax-eq} takes the following
form of partial differential equations
\be
\pa_r u_n = F_{n,r} \( u_0, u_1,\ldots, u_{r+n-1} \)\qquad n= 0, \ldots
\lab{pde}
\ee
with differential polynomials $F_{n,r}$ in $u_0, u_1,\ldots, u_{r+n-1}$.
In particular we have a hierarchy of flow equations:
\be
\pa_r u_0 \equiv F_{r} \( u_0, u_1,\ldots, u_{r-1} \)
\lab{hflows}
\ee
Equations \rf{hflows} can be viewed as symmetries of the
KP equation according to our above definition.

We will first introduce a constrained KP (\cKP) hierarchy by imposing a
constraint on symmetry represented by the evolution equations in \rf{hflows}.

We will discuss three (very related) approaches to define the \cKP hierarchy
and show their equivalence.
\mskp
{\bf Symmetry constraint, \cKP hierarchy, Definition I}

This approach defines the constrained KP hierarchy by imposing the
symmetry constraint on flows in \rf{hflows} in terms of
the eigenfunction and its adjoint entering the linear
problems \rf{ba-b} and \rf{staba-b}:
\be
F_{r} \( u_0, u_1,\ldots, u_{r-1} \) = \pa_x (\Phi \Psi)
\lab{symcons}
\ee
Hence for each $r=1,2,\ldots$ we define the different \cKP hierarchies
by imposing equality of
the flows/symmetries $F_r$ to the symmetry $\pa_x (\Phi \Psi) $ of the KP
equation.
The constraint allows us to eliminate the Lax coefficients $u_{k\geq r-1}$
in terms of eigenfunctions $\Phi$ and $ \Psi$ and $u_{k<r-1}$.

The next two approaches will allow us to write down an explicit
Lax representation of the \cKP hierarchy.
\mskp
{\bf Symmetry constraint, \cKP hierarchy, Definition II}
\mskp
Consider the Lax operator
$Q=D+ \sumi{i=0} u_i D^{-i-1}$ satisfying the flow equations
as in \rf{lax-eq} i.e.
\be
\partder{Q}{t_k} = \lb B_k\, , \, Q\rb
\quad ;\quad   B_k \equiv (Q^k)_{+}
\quad  \; \; k= 1, 2, \ldots \lab{lax-eqq}
\ee
Let furthermore (like in \rf{eigenfcts}) $\P$ and $\Psi$ be, respectively,
eigenfunction and adjoint eigenfunction of $Q$, meaning that for
$k= 1, 2, \ldots$ it holds
\be
\partder{\Phi}{t_k} = B_k \Phi \qquad ; \qquad
\frac{\pa \Psi}{\pa t_k} = - B_k^{*} \Psi
\lab{adeigen}
\ee
where $B_k^{*} $ is an adjoint operator of $B_k$.
We now impose the constraint on the KP hierarchy by requiring that
the Lax operator $Q$ satisfies a condition \ct{cheng}:
\be
Q^r = B_r + \Phi D^{-1} \Psi  \quad;\quad B_r \equiv (Q^r)_{+}
\lab{cheng}
\ee
for $r$ being a fixed positive integer.

Let us first verify that {\sl Definition II} implies the symmetry constraint
of {\sl Definition I}.
Let us namely impose constraint \rf{cheng} and notice that from
\rf{lax-eqq} we get
\br
\pa u_0 / \pa t_r &=& {\rm Res}_{\pa} \( \pa Q/ \pa t_r\)
= {\rm Res}_{\pa} \lb B_r\, , \, Q \rb
= {\rm Res}_{\pa} \lb Q^r - \Phi D^{-1} \Psi \, , \, Q \rb \lab{cheko} \\
&=& {\rm Res}_{\pa} \lb - \Phi D^{-1} \Psi \, , \, Q \rb = \pa_x \(\Phi\Psi\)
\nonu
\er
This clearly shows that the {\sl Definition II} realizes directly in the
Lax setting the symmetry constraint of {\sl Definition I}.

\lemma {\em The time evolution of the pseudo-differential operator
$ \Phi D^{-1} \Psi$ is qiven by:}
\be
\partder{}{t_k} \Phi D^{-1} \Psi = \lb B_k\, , \, \Phi D^{-1} \Psi\rb_{-}
\lab{tkppsi}
\ee
The proof is a consequence of following technical observations based on
\rf{adeigen}:
\be
\(B_k \Phi D^{-1} \Psi\)_{-} = \(B_k \Phi\)_{0} D^{-1} \Psi =
\partder{\Phi}{t_k} D^{-1} \Psi
\lab{bkppsi}
\ee
and
\be
-\(\Phi D^{-1} \Psi B_k \)_{-} = -  \Phi D^{-1} \(B_k^{\ast} \Psi\)_{0} =
\Phi D^{-1} \partder{\Psi}{t_k}
\lab{ppsibk}
\ee

What we learn from the above lemma is that if we define  the Lax operator
$L$ such that its purely pseudo-differential part is $L_{-} =
\Phi D^{-1} \Psi$ then $L_{-} $ satisfies automatically the KP flow equations:
$\pa L_{-}/ \pa t_k = \lb B_k\, , \, L \rb_{-}$.
Note that in the last equation we used that $ \lb B_k\, , \, L \rb_{-}=
\lb B_k\, , \, L_{-} \rb_{-}$.
These observations lead us to  the next definition of \cKP hierarchy.
\mskp
{\bf Symmetry constraint, \cKP hierarchy, Definition III}
\mskp
Here we work with Lax operator of the \cKP hierarchy defined as
\br
L &=& D^r+ \sum_{l=0}^{r-2} U_l D^l + \Phi D^{-1} \Psi
\lab{f-5}  \\
\partder{ L}{t_k} &=& \lb \( L^{k/r} \)_{+} \, , \, L \rb
\lab{f-5a}
\er
We first address an issue of equivalence between {\sl Definition II} and
{\sl Definition III}, which is easy to establish
when making a connection $ Q^r \equiv L$.
It is then easy to see that the two flows given below are equivalent
($B_k = \(Q^k\)_{+}=\(L^{k/r}\)_{+}$):
\be
\pa Q / \pa t_k = \llb\(Q^k\)_{+}  \, , \, Q\rrb \; \leftrightarrow \;
\pa L / \pa t_k = \llb\(Q^k\)_{+}  \, , \, L \rrb =
\llb\(L^{k/r}\)_{+}  \, , \, L\rrb
\lab{equi-23}
\ee
We have already seen in \rf{tkppsi} that the pseudo-differential part of
$L$ from \rf{f-5} evolves according to the Sato's formalism.
Let us now investigate evolution of the differential part $B_r = \(Q^r\)_{+}
=\(L\)_{+}$ of $L$:
\be
\pa B_r  / \pa t_k  = \lb B_k  \, , \, L\rb_{+}=
\lb B_k  \, , \, B_r \rb + \lb B_k  \, , \, \Phi D^{-1} \Psi\rb_{+}
\lab{ff-5}
\ee
which after using Zakharov-Shabat equation \rf{zs-eq} becomes:
\be
\pa B_k / \pa t_r = \lb B_k  \, , \, \Phi D^{-1} \Psi\rb_{+}
\lab{ff-6}
\ee
valid for all $k=1,2, \ldots\;$.
Equation \rf{ff-6} describes how the \cKP constraint condition is being imposed
on the flows of KP coefficients in $B_k$.
Let us study consequences of \rf{ff-6} in some simple cases.
The first nontrivial case occurs for $k=2$ with $B_2 =\(Q^2\)_{+}= D^2
+ 2 u_0 = D^2 + U_0 $. From \rf{ff-6} it follows:
\be
2 \pa u_0 / \pa t_r = \lb D^2  \, , \, \Phi D^{-1} \Psi\rb_{+}
= 2 \pa_x (\Phi \Psi)
\lab{repro1}
\ee
which reproduces the flow equation being basis for the {\sl Definition I}
of the \cKP hierarchy, establishing henceforth equivalence between
{\sl Definition I} and {\sl Definition III}.

For $k=3$ we have $B_3=\(Q^3\)_{+}= D^3 + 3 u_0 D + 3 u_1 + 3 \pa_x u_0
= D^3 + U_0 D + U_1 $.
{}From \rf{ff-6} it follows this time that:
\be
\pa (3 u_0 D + 3 u_1 + 3 u_0^{\pr}) / \pa t_r =
\lb D^3\, , \, \Phi D^{-1} \Psi\rb_{+}
= 3 \Phi^{\pr \pr} \Psi + 3 \Phi^{\pr } \Psi^{\pr} +
3 \(\Phi \Psi\)^{\pr} D
\lab{repro2}
\ee
or
\be
\pa u_1 / \pa t_r = - \(\Phi \Psi^{\pr} \)^{\pr}
\lab{repro3}
\ee
Consider the Lax operator $L$ having the form as in \rf{f-5} and
satisfying \rf{f-5a}.
One can ask whether $\Phi$ ($\Psi$) are automatically the (adjoint)
eigenfunctions.
The following Lemma answers this question affirmatively.

{\bf Lemma.}\quad {\sl Let $L = L_{+} +  \Phi D^{-1} \Psi$,
with $\, r\,$ being the order of $L_{+}$. Then the Lax equations of motion
$\, \partder{}{t_k } L = \sbr{B_k }{L}\,$ imply:
\be
\llb \( \partder{}{t_k}\Phi- \( B_k \Phi\)_0\)
D^{-1} \Psi+ \Phi D^{-1} \( \partder{}{t_k}\Psi +
\( B_k^{\ast} \Psi \)_0 \) \rrb = 0   \lab{f-2}
\ee
where $B_k = (L^{k/r})_{+}\;, \;B_k^{\ast} = (L^{k/r})_{+}^{\ast}$}
\lskip
\mark Eq. \rf{f-2} is equivalent to the infinite set of equations:
\be
\llb \( \partder{}{t_k}\Phi- \( B_k\Phi\)_0\)
\pa_x^l \Psi+ \Phi\pa_x^l \( \partder{}{t_k}\Psi+
\(B^\ast_{k}\Psi\)_0 \) \rrb = 0  \quad ,\quad
l=0,1,2,\ldots  \lab{f-3}
\ee
An obvious solution of \rf{f-3} seems to be:
\br
\partder{}{t_k}\Phi- \( B_{k} \Phi\)_0 =
\partder{c_r}{t_k}\, \Phi\lab{f-4} \\
\partder{}{t_k}\Psi+ \( B^{\ast}_{k} \Psi\)_0 =
- \partder{c_r}{t_k}\, \Psi\nonu
\er
where $c_r$ is $x$-independent. Thus, up to a $x$-independent phase
transformation $ \Phi\to e^{c_r}\Phi$ and $ \Psi\to e^{-c_r}\Psi$,
which does not change the Lax operator, $\Phi$ ($\Psi$) are (adjoint)
eigenfunctions.

\mskp
\underbar{Examples:}

{\bf r=1}. The simplest case is when $r = 1$ for which we have:
\be
F_1 = \pa_x u_0 = \pa_x \Phi \Psi \to  u_0 = \Phi \Psi + {\rm const}
\lab{monecase}
\ee
Inserting this into \rf{eigentwo} (with zero integration constant) we get
\be
\partder{\Phi}{t_2} = {\pa^2 \Phi \o \pa x^2} + 2 \Phi^2 \Psi\quad;\quad
\partder{\Psi}{t_2} = -{\pa^2 \Psi \o \pa x^2} -
2 \Phi \Psi^2
\lab{eigenonered}
\ee
in which we recognize the Nonlinear Schr\"{o}dinger (NLS) equations
being the first (non-trivial) flow of the AKNS hierarchy.
Eqs. \rf{eigenonered} are also results of the Sato equation \rf{lax-eq}
for two-boson hierarchy \ct{BAK85,2boson} defined by the Lax operator
$Q=L= D + \Phi D^{-1} \Psi$. Expressions for the coefficients $u_{k \geq 0}$
of the standard expansion for $Q$ are easy to find in terms of the \faa
polynomials \ct{2boson,swieca-z}.

{\bf r=2}. Here $F_2= \pa_x^2 u_0 + 2 \pa_x u_1= \pa_x \( \Phi \Psi\) $
which allows to express $u_{n \geq 1}$ by $u_0, \Phi, \Psi$.
Alternatively writing in the spirit of the {\sl Definition III}
the Lax operator as $L=Q^2= D^2 + 2 u_0 + \Phi D^{-1} \Psi$ we obtain from the
Sato eq. \rf{equi-23} or \rf{eigenfcts}:
\br
\partder{\Phi}{t_2} &=& {\pa^2 \Phi \o \pa x^2} + 2 u_0 \Phi
\nonu \\
\partder{\Psi}{t_2} &=& -{\pa^2 \Psi \o \pa x^2} -2 u_0 \Psi \lab{twored}\\
\partder{u_0}{t_2} &=& \pa_x (\Phi \Psi)
\nonu
\er
which agrees with first flow equation of the so-called Yajima-Oikawa hierarchy
\ct{yajima,konop}

{\bf r=3}. Here $F_3 = \pa_x^3 u_0 + 3 \pa_x^2 u_1 +3 \pa_x u_2
+ 6 u_0 \pa_x u_0 = \pa_x \( \Phi \Psi\) $
which allows to express $u_{n \geq 2 }$ by $u_0,u_1, \Phi, \Psi$.
Alternatively, writing in the spirit of the {\sl Definition III}
the Lax operator as $L=Q^3 = D^3 + 3 u_0 D + 3 u_1 + 3 \pa_x u_0+
\Phi D^{-1} \Psi$ we obtain from the
Sato eq. \rf{equi-23}:
\br
\partder{\Phi}{t_2} &=& {\pa^2 \Phi \o \pa x^2} + 2 u_0 \Phi
\nonu \\
\partder{\Psi}{t_2} &=& -{\pa^2 \Psi \o \pa x^2} -2 u_0 \Psi \nonu \\
\partder{u_0}{t_2} &=& \pa_x^2 u_0 +2  \pa_x u_1 \lab{threered}\\
\partder{u_1}{t_2} &=& - \pa_x^2 u_1 -{ 2\o 3}  \pa_x^3 u_0
-2 u_0 \pa_x u_0 +{ 2\o 3} \pa_x \( \Phi \Psi\) \nonu
\er
in which one recognize the $t_2$ flow of the so-called
Melnikov system \ct{melnikov}.
The flow eqs. for $t_3=t_r$ follow from \rf{repro1}, \rf{repro2} and
\rf{eigenfcts} (or \rf{eigenthree}):
\br
\partder{\Phi}{t_3} &=& {\pa^3 \Phi \o \pa x^3} + 3 u_0 {\pa \Phi \o \pa x}
+ 3 \( u_1 + {\pa u_0 \o \pa x }\) \Phi \nonu \\
\partder{\Psi}{t_3} &=& {\pa^3 \Psi \o \pa x^3} +3 u_0 {\pa \Psi \o \pa x}
-3 u_1 \Psi \nonu \\
\partder{u_0}{t_3} &=&\pa_x (\Phi \Psi)  \lab{threereda}\\
\partder{u_1}{t_3} &=& - \(\Phi \Psi^{\pr} \)^{\pr}  \nonu
\er
\mskp
In the next subsection we will learn how to view the above examples as special
cases of $SL (r+1,1)$ \cKP hierarchy with $r=1,2,3$ respectively.

\mark In the above we have kept for simplicity discussion restricted
to a single pair $\P, \Psi$ of (adjoint) eigenfunctions.
It is very simple to generalize all formulas to the case
of $n$ $\Phi_n, \Psi_n$ (adjoint) eigenfunctions.
Clearly in expressions \rf{cheng} and \rf{f-5} we need to make
a substitution: $ \Phi D^{-1} \Psi \to \sum_{i=1}^n \Phi_i D^{-1} \Psi_i$.
This is the structure, which will arise naturally in the next subsection.

\subsection{\cKP Hierarchy as ${\bf SL(m,n)}$-type Lax Hierarchy
and its Bi-Poisson Structure }
Let us introduce a ratio of two differential operators (with $m =r+n$
in notation of the previous subsection).
\be
\cL_{m,n} \equiv { L^{(m)} \o  L^{(n)} }  \qquad \; n\leq m-1
\lab{ratio-a}
\ee
where
\be
L^{(m)} = D^m + u_{m-1} D^{m-1} + \ldots + u_1 D + u_0
\quad ;\quad
L^{(n)} = D^n + \tu_{n-1} D^{n-1} + \ldots + \tu_1 D + \tu_0
\lab{lmk}
\ee
Let $\{ \psi_i^{(m)}\; ,\; i =1, \ldots,m \}$ and $\{ \psi_i^{(n)}
\; ,\; i =1, \ldots,n \}$ be a basis for the kernels of $L^{(m)}$
and $L^{(n)}$, respectively, i.e. we have $L^{(\a)} \psi_i^{(\a)}=0\;,\;
i =1, \ldots,\a=m \,\,{\rm or}\,\,n$.
Alternatively, we can rewrite \rf{lmk} as
\be
L^{(m)} = \( D + v_m\) \( D + v_{m-1} \) \cdots \( D + v_1 \)
\quad ;\quad
L^{(n)} = \( D + \tv_{n} \) \( D + \tv_{n-1} \) \cdots \( D + \tv_1 \)
\lab{lmk-a}
\ee
with
\be
v_i = \pa \( \log{ W_{i-1} \lb \psi_1^{(m)}, \ldots, \psi_{i-1}^{(m)} \rb \o
W_{i}\lb \psi_1^{(m)}, \ldots, \psi_{i}^{(m)} \rb  }  \) \qquad, \qquad
W_0 =1
\lab{wil}
\ee
and similar expressions for $\tv_i$.

Taking the above into consideration we can rewrite \rf{ratio-a} as
\ct{office,yu}
\be
\cL_{m,n} = \prod_{j=m}^{1} \( D + v_j \) \prod_{l=1}^{n} \( D + \tv_l\)^{-1}
\lab{ratio-b}
\ee
(see also \ct{BX9305,dickey94,oevels} for discussion of the
pseudo-differential operators of the constrained KP hierarchy).
We will now study the Poisson structures of the hierarchy defined by
the Lax operators of the form given in \rf{ratio-b}.
It is easy to  prove the following Proposition \ct{yu,office}.

\prop
{\em The Poisson bracket relation:}
\br
\{ v_i \, ,\, v_j \}_{PB} \eq  \d_{ij} \; \d^{\pr} (x-y)
\quad  \; \;i,j =1,\ldots m\nonu\\
\{ \tv_r \, ,\, \tv_s\}_{PB} \eq -\d_{rs} \; \d^{\pr} (x-y) \quad  \; \; \;
r,s = 1 ,\ldots , n  \lab{pbbb}\\
\{ v_i \, ,\, \tv_l\}_{PB} \eq  0
\nonu
\er
{\em is equivalent to}
\be
{\pbbr{\me{\cL_{m,n}}{X}}{\me{\cL_{m,n}}{Y}}}_{PB} =
{\Tr}_A \( \( \cL_{m,n}X\)_{+} \cL_{m,n}Y -
\( X\cL_{m,n}\)_{+} Y\cL_{m,n} \)
\lab{lprop}
\ee
where the subscript $PB$ stands for the Poisson bracket as defined by
\rf{pbbb}. Moreover the Adler trace $ \me{\cL_{m,n}}{X}$ defines an invariant,
non-degenerate bilinear form: $\int dx {\rm Res} \(\cL_{m,n} X\)$
for the pseudo-differential operator $\cL_{m,n}$ and its dual $X$
(see for instance \ct{swieca-z,office}).

We now introduce a Dirac constraint:
\be
\Psi_{m,n} = \sum_{j=1}^{m} v_j - \sum_{l=1}^{n}  \tv_l= 0
\lab{psimk1}
\ee
which is  second-class due to:
\be
\{ \Psi_{m,n}\, , \, \Psi_{m,n}\}_{PB} = (m-n)  \;\d^{\pr} (x-y)
\lab{psimkbra}
\ee
Condition \rf{psimk1} expresses tracelessness of the underlying graded
$SL (m,n)$ algebra.
Corresponding Poisson algebra of diagonal part of the graded
$SL (m, n)$ Kac-Moody algebra is obtained by the usual Dirac bracket
calculation:
\br
\{ v_i \, ,\, v_j \}_{DB} \eq  \(\d_{ij} -  {1 \o m-n}\) \d^{\pr} (x-y)
\quad\quad i,j =1,\ldots, m\nonu\\
\{ {\ti v}_r \, ,\, {\ti v}_s \}_{DB} \eq -\(\d_{rs} + {1 \o m-n }\) \d^{\pr}
(x-y) \quad\quad  r,s = 1, \ldots,n  \lab{redbbb}\\
\{ v_i \, ,\, {\ti v}_r \}_{DB} \eq  {1 \o m-n}\, \d^{\pr} (x-y)
\nonu
\er

\prop
{\em The Dirac bracket \rf{redbbb} takes the following form in the Lax operator
representation:}
\br
{\pbbr{\me{\cL_{m,n}}{X}}{\me{\cL_{m,n}}{Y}}}_{DB} \eq
{\pbbr{\me{\cL_{m,n}}{X}}{\me{\cL_{m,n}}{Y}}}_{PB} \lab{ldb}\nonu\\
&+&{1 \o m-n} \int dx \; {\rm Res}\( \sbr{\cL_{m,n}}{X}\) \pa^{-1}
{\rm Res}\( \sbr{\cL_{m,n}}{Y}\)
\nonu
\er
The proof follows from evaluation of the extra term of the relevant Dirac
bracket:
\be
-\int {\pbbr{\me{\cL_{m,n}}{X}}{\Psi_{m,n}}}_{PB}
\;{\pbbr{\Psi_{m,n}}{\Psi_{m,n}}}_{DB}^{-1}\;
{\pbbr{\Psi_{m,n}}{\me{\cL_{m,n}}{Y}}}_{DB}
\ee
One easily verifies \rf{ldb} using \rf{psimkbra}:
\be
 {\pbbr{\me{\cL_{m,n}}{X}}{\Psi_{m,n}(z)}}_{PB} =
- \me{\sbr{\d (x-z)}{\cL_{m,n}}}{X}
=-{\rm Res}\( \sbr{X}{\cL_{m,n}}\)(z)
\ee

Formula \rf{ldb} contains as special cases $n=m-1$ corresponding to the
constrained KP hierarchy with $L_+=D$ and $n=0$ corresponding to the
$m$-KdV hierarchy (see e.g. \ct{diz}).
For the intermediary cases $ 0 < n < m-1$ this formula presents a compact
expression for the bracket structure of the $SL (m,n)$ \cKP hierarchy.

The Lax operator from \rf{ratio-a} restricted to the constrained manifold
defined by \rf{psimk1} can be parametrized as
\be
L_{m,n} \equiv \cL_{m,n}\bv_{\Psi_{m,n}=0} =
\prod_{l=m-1}^{n+1} \( D + {\wti c}_{l}\)
\prod_{l=n}^{1} \( D +  {\wti c}_{l}+ {\ti v}_{l}\)
\( D - \sum_{l=1}^{m-1} {\wti c}_{l} \)\prod_{l=1}^{n}
\( D + {\ti v}_{l}\)^{-1}
\lab{laxmk}
\ee
We can alternatively rewrite the expression \rf{laxmk} as
\be
L_{m,n}
= \sum_{l=1}^{n} {\wti A}_l  \prod_{i=l}^{n} \( D + {\tv}_{i}\)^{-1}
+ \sum_{l=0}^{m-n-2} {\wti A}_{l+n+1}  D^{l} + D^{m-n} \lab{1ff}
\ee
with the second bracket structure automatically given by the formula \rf{ldb}.

For the special case $m-n=r=1$ we have a canonical representation for
variables $\(v_i, \tv_l\)$ with $ 1 \leq i \leq m \; ,\; 1 \leq l\leq m-1$
in
$L_{m,m-1} = \prod_{j=m}^{1} \( D + v_j \) \prod_{l=1}^{m-1}
\( D + \tv_l\)^{-1}$
in terms of the pairs $(c_r ,e_r )_{r=1}^{m-1}\,$, which are
the ``Darboux'' canonical pairs for the second KP bracket satisfying
$\{ c_i (x) \, ,\, e_l(y) \}= - \d_{il} \pa_x \d (x-y)$ for
$i,l =1,2, \ldots, m-1$.
The representation is \ct{office}:
\br
\tv_l &=& - e_l - \sum_{p=l}^{m-1} c_p    \qquad l =1,2, \ldots, m-1
\lab{tvec}\\
v_i &=& - e_{i-1} - \sum_{p=i}^{m-1} c_p= -c_i + \tv_i  \qquad i=1, \ldots, m
\lab{vec}
\er
Eq. \rf{vec} includes the special cases of $v_1 = - \sum_{p=1}^{m-1} c_p$
and $v_m = -e_1$.
One checks easily that in this representation the Poisson algebra
of $\tv_l, v_i$ takes indeed the form of the bracket algebra
of graded $SL (m,m-1)$ Kac-Moody algebra in a diagonal gauge:
\br
\{ v_i \, ,\, v_j \} \eq  \(\d_{ij} -  1\) \, \d^{\pr} (x-y) \quad
\quad i,j =1,\ldots, m\nonu\\
\{ \tv_p \, ,\, \tv_l \} \eq -\(\d_{pl} + 1 \) \d^{\pr} (x-y) \quad \quad
p,l=1, \ldots,m-1  \lab{bbbij}\\
\{ v_i \, ,\, \tv_l \}\eq  -\d^{\pr} (x-y)
\nonu
\er

We can interpret \rf{laxmk} as a superdeterminant of the graded $SL (m, n)$
matrix in a diagonal gauge, which for KdV case $n=0$ becomes an ordinary
determinant as in Fateev-Lukyanov \ct{FL88} expression:
\be
L_{m,0} = \prod_{i=1}^{m} \( D  + v_{i} \)
= D^{m} + {\wti A}_{m-1} D^{m-2} + \ldots + {\wti A}_1 \quad ;
\quad \sum_{i=1}^{m} v_i =0
\lab{fateev}
\ee

\underbar{Example}. As an example let us take $m=2, n=1$ in \rf{laxmk}. Then:
\be
L_{2,1}=  \( D  + c_1 + \tv_1 \) \( D  - c_1 \)
\( D  + \tv_1 \)^{-1} = - \llb(\tv_1 + c_1)^{\pr} +  c_1
(\tv_1 + c_1) \rrb \( D  + \tv_1 \)^{-1} +D
\lab{mtwokone}
\ee
as follows by direct calculation making use of an obvious identity
$( D  + c)\,( D  + v )^{-1} = (c-v) \,( D  + v )^{-1} +1$.

One notices the absence of the term proportional to $D^{m-n-1}$ in \rf{1ff}
due to the $SL (m,n)$ trace zero condition.
Another feature is the presence of the constant term $A_{n+1}$ in
the Lax operator $L_{m,n}$ (for $m-n >1$).
This fact enables us to prove that $SL (m, n)$-\cKP hierarchy is a
bi-Poisson hierarchy. Consider namely $L_{m,n}^{\pr}=
L_{m,n}- \l $ obtained by redefining the $D^0=1$ term in the Lax
operator by addition of the constant $\l$.
Clearly the value of the of bracket \rf{ldb} for the new Lax is
\br
&&\!\!\!{\pbbr{\me{L_{m,n}^{\pr}}{X}}{\me{L_{m,n}^{\pr}}{Y}}}_{DB}
\! = {\Tr}_A \( \(L_{m,n}^{\pr}X\)_{+} L_{m,n}^{\pr}Y-
\( X L_{m,n}^{\pr}\)_{+} Y L_{m,n}^{\pr}\)\nonu\\
&+&{1 \o m-n} \int dx  {\rm Res}\( \sbr{L_{m,n}^{\pr}}{X}\) \pa^{-1}
{\rm Res}\( \sbr{L_{m,n}^{\pr}}{Y}\)
- \l \llangle L_{m,n}^{\pr}\bv \left\lb X,\, Y \right\rb_{R}\rrangle
\lab{pencil}
\er
where we introduced an $R$-commutator $\lb X,\, Y \rb_{R} \equiv
\lb X_{+},\, Y_{+} \rb-\lb X_{-},\, Y_{-}\rb$
with subscripts $\pm$ denoting projection on pure differential and
pseudo-differential parts of the pseudo-differential operators $X,Y$.
Define next an $R$-bracket $ \{\cdot ,\cdot \}_{1}^{R}$
as a bracket obtained by substituting
$R$-commutator $\lb X,\, Y \rb_{R}$ for the ordinary commutator
\ct{STS83,R82,ANPV}:
\be
{\pbbr{\me{L}{X}}{\me{L}{Y}}}_1^{R} \equiv
- \llangle L \bv \left\lb X,\, Y \right\rb_{R} \rrangle  \lab{first-RKP}
\ee
Relation \rf{pencil} shows that $\{\cdot ,\cdot \}_{DB} +
\l \{\cdot ,\cdot \}_{1}^{R}$ satisfies the Jacobi identity.
We can state this result as:

\prop
{\em ${SL (m, n)}$-\cKP hierarchy is \underbar{bi-Poisson} with brackets
$\{\cdot ,\cdot \}_{DB}$ and $\{\cdot ,\cdot \}_{1}^{R}$ defining a compatible
pair of brackets.}

This Proposition establishes the fundamental criterion for
integrability of the ${SL (m, n)}$-\cKP hierarchy.
Clearly the argument holds also for the case $m-n=1$, where
one adds the constant $\l$ to zero representing
the missing constant term.

Let us now concentrate on the pseudo-differential part of
$L_{m,n}$ from \rf{1ff}.
We can rewrite it as
\br
\(L_{m,n}\)_{-} &= &\sum_{l=1}^n r_l \prod_{i=l}^n D^{-1} q_i
\lab{iss-8aaa} \\
r_l= {\wti A}_{l} e^{-\int \tv_{l}} \quad ;\;\; q_n= e^{\int \tv_n} \;\; &,&
\;\; q_i= e^{\int \( \tv_{i} - \tv_{i+1}\)} \;\;, \;\;
i=1,\ldots ,n-1  \lab{4.5}
\er
Let us define the quantity
\be
Q_{l,i} \equiv (-1)^{i-n} \int q_i \int q_{i-1} \int \ldots
\int q_l\, (dx^{\pr})^{i-l+1}
\quad \qquad 1 \leq l\leq i \leq n
\lab{qni}
\ee
Then using that $ D^{-1} Q_{1,i-1}  q_i = D^{-1} Q_{1,i} D-
Q_{1,i}$ we obtain for quantity in eq.\rf{iss-8aaa}
\be
\(L_{m,n}\)_{-}= \sum_{i=2}^{n}  r_i^{(1)} \prod_{l=i}^n D^{-1} q_l
+ r_1 D^{-1} \( - Q_{1,n-1} q_n\)
\lab{rqlax-a}
\ee
where
\be
r_i^{(1)} \equiv r_i +r_1 Q_{1,i-1} \qquad i=2, \ldots , n
\lab{ronei}
\ee
The above process can be continued to yield an expression
\be
\(L_{m,n}\)_{-}=  \sum_{i=1}^n \Phi_i D^{-1} \Psi_i
\lab{f-5b}
\ee
with
\br
\P_i \eq r_i + \sum_{n=1}^{i-1} r_{n}
\sum_{s_{i-n-1} =s_{i-n-2} +1}^{i-n} \cdots \sum_{s_2 =s_{1} +1}^{i-n}
\sum_{s_{1} = 1}^{i-n} Q_{n,i-s_{i-n-1}-1}
Q_{i-s_{i-n-1},i-s_{i-n-2}-1} \cdots                    \nonu\\
&\cdots& Q_{i-s_2,i-s_{1}-1} Q_{i-s_{1},i-1}
\qquad \qquad\qquad \qquad 1 \leq i \leq n
\lab{piri} \\
\Psi_n \eq q_n \quad\; , \quad \;
\Psi_i = (-1)^{n-i} q_n \int q_{n-1} \int \ldots \int q_i
\, (dx^{\pr})^{n-i}  \qquad 1\leq i \leq n-1
\lab{psiqi}
\er
Note that the new variables $\Psi_i$ coincide with elements $\psi^{(n)}_i$
in the kernel of $L^{(n)}$.
It follows from the construction opposite to the one shown above and involving
the following relation (see \ct{avoda} for more details):
\br
\(L_{m,n}\)_{-}&=&  \sum_{i=1}^n \Phi_i D^{-1} \Psi_i
\lab{iss-8aa}  \\
&=&  \sum_{i=1}^n A^{(n)}_i \( D + B^{(n)}_i \)^{-1}
\( D + B^{(n)}_{i+1}\)^{-1} \cdots \( D + B^{(n)}_n \)^{-1}   \lab{iss-8a}
\er
where the new variables are :
\br
A^{(n)}_i&=& (-1)^{n-i} \sum_{s=1}^i \Phi_s \frac{W\llb \Psi_n ,\ldots ,
\Psi_{i+1},\Psi_s \rrb}{W\llb \Psi_n,\ldots ,\Psi_{i+1}\rrb} \lab{iss-8b}\\
B^{(n)}_i &=& \pa_x \ln \frac{W\llb \Psi_n ,\ldots ,
\Psi_{i+1},\Psi_i\rrb}{W\llb \Psi_n,\ldots ,\Psi_{i+1}\rrb} \lab{iss-8c}
\er
{}From \rf{iss-8a} we see that ${\ti v}_i = B^{(n)}_i $ and  from
\rf{iss-8c} it follows that $\psi^{(n)}_i = \Psi_i$.

\underbar{Example}. For $n=2$ relations \rf{iss-8b}-\rf{iss-8c} become
\br
A^{(2)}_2 &=& \Phi_1 \Psi_1 + \Phi_2 \Psi_2  \quad , \quad
A^{(2)}_1 = - \Phi_1 \Psi_1 \( \pa_x \ln \Psi_2^{-1} \Psi_1 \)
  \lab{iss-2}  \\
B^{(2)}_2 &=& \pa_x \ln \Psi_2  \quad , \quad
B^{(2)}_1 = \pa_x \ln \llb \Psi_1 \( \pa_x \ln \Psi_2^{-1} \Psi_1 \) \rrb
  \lab{iss-2a}
\er
Equivalence between \rf{iss-8a} and \rf{iss-8aa} follows then
by inspection.

\prop $\Phi_i, \Psi_i$ are canonical-Darboux fields for the
first bracket of the $SL (n+1,n)$ \cKP hierarchy :
\be
{\pbr{\Phi_i}{\Psi_j}}_1 = - \d_{ij} \d (x-y) \qquad,\qquad i,j = 1,
\ldots,n
\lab{phipsi-can}
\ee

\subsection{Affine ${\bf sl (n+1)}$ Origin  of ${\bf SL (n+1,n)}$
\cKP Hierarchy. }

In this subsection we will establish a connection between Generalized
Non-linear Schroedinger $\GNLS$
matrix hierarchy for the hermitian symmetric space $sl (n+1)$ \ct{FK83} and
the constrained KP hierarchy.

We first introduce ZS-AKNS scheme arising from
the linear matrix problem for $A \in \lie$ with $\lie$ being a Lie algebra
\ct{FNR83,ne85}:
\br
\pa \Psi \eq  A \Psi  \qquad, \qquad \pa = \partder{}{t_1}= \pa_x \lab{1.1} \\
\pa_{t_m} \Psi \eq B_m \Psi \qquad m=2,3,\ldots \lab{1.2}
\er
The compatibility condition for the linear problem \rf{1.1} and \rf{1.2}
leads to the Zakharov-Shabat (Z-S) integrability equations
\be
\pa_{m} A - \pa B_m + \lb A \, , \, B_m \rb = 0 \quad;\quad
\pa_m \equiv  \pa_{t_m}
\lab{1.6}
\ee
Let us use the following decomposition in \rf{1.1}:
\be
A = \l E + A^{0} \qquad {\rm with} \qquad
E = { 2 \mu_a \cdot H \o \a_a^2}
\lab{1.3}
\ee
where $\mu_a$ is a fundamental weight and $\a_a$ are
simple roots of $\lie$.
The element $E$ is used to decompose the Lie algebra
$\lie$ as follows:
\be
\lie = {\rm Ker} ( {\rm ad} E) \oplus {\rm Im } ( {\rm ad} E)
\lab{1.5}
\ee
where the ${\rm Ker} ( {\rm ad} E)$  has the form $
\cK ^{\pr} \times u(1)$ and is spanned by
the Cartan subalgebra of $\lie$ and step operators associated to roots not
containing $\a_a$.
Moreover ${\rm Im } ( {\rm ad} E)$ is the orthogonal complement of
${\rm Ker} ( {\rm ad} E)$.

{}From now on we consider $\lie = sl(n+1)$ with roots $\a = \a_i +
 \a_{i+1} +\ldots + \a_j $ for some $i,j=1,\ldots ,n$ , $E = {2\mu_n.H \o
{\a_n^2}}$, $\mu_n$ is the $n^{\rm th}$ fundamental weight and $H_a$, $a=1,
\ldots , n$ are the generators of the Cartan subalgebra. This
decomposition  generates the symmetric space $sl(n+1)/sl(n)\times u(1) $
(see \ct{affine} for details).

ZS-AKNS scheme becomes in this case the $\GNLS_n$
(=$sl (n+1)$ GNLS) hierarchy.
In matrix notation we have:
\be
E = {1 \o n+1} \left(\begin{array}{ccccc}
1 & & & & \\  & 1& & & \\  & &1 & & \\ & & &\ddots &\\
& & & & -n \end{array} \right)
\lab{1.4}
\ee
The model is defined by $A^{0} \in {\rm Im} ( {\rm ad} E)$
and is parametrized by fields $q_a$ and $r_a$, $a=1,\ldots ,n$
according to:
\be
A^{0}= \sum_{a=1}^n \( q_a E_{(\a_a+ \ldots+\a_n)}+ r_a E_{-(\a_a+ \ldots
+\a_n )}  \)
\lab{1.11}
\ee
which in the matrix form can be rewritten as
\be
A^{0} = \left(\begin{array}{ccccc}
0 &\cdots &0 &\cdots & q_1 \\
0 &0 &\cdots &0 &q_2 \\
0& & \ddots & & \vdots \\
\vdots& &  &\ddots & q_n \\
r_1 &r_2  &\cdots & r_n & 0
\end{array} \right)
\lab{1.12}
\ee
It can be shown that the flows as defined in \rf{1.1} and \rf{1.3} satisfy
the recurrence relation:
\be
\pa_m A^{0} = \cR \pa_{m-1} A^{0} \quad;\quad
\cR \equiv  \( \pa - ad_{A^0}\, \pa^{-1} ad_{A^0} \) ad_{E}
\lab{1.18}
\ee
where we have defined a recursion operator $\cR $ \ct{FK83,affine}.

The connection to the \cKP hierarchy is first established between linear
systems defining both hierarchies.
With \rf{1.4} and \rf{1.12} the linear problem from \rf{1.1} is explicitly
given by:
\be
\left(\begin{array}{ccccc}
\pa-\l/(n+1) &0 &\cdots &0 &q_1 \\
0 & \pa-\l/(n+1) &0 &\cdots &q_2 \\
\vdots & &\ddots & &\vdots \\ 0 & & &\pa-\l/(n+1)  & q_n\\
r_1& r_2 &\cdots &r_n & \pa+n\l/(n+1)  \end{array} \right)
\left(\begin{array}{l}
\psi_1 \\ \psi_2 \\ \vdots \\  \psi_n \\ \psi_{n+1}
\end{array} \right)   =        0
\lab{3.1}
\ee

Perform now the phase transformation:
\be
\psi_k \; \longrightarrow \; {\bar \psi_k} = \exp \( - {1 \o n+1} \int
\l\, dx \) \psi_k \qquad \; k = 1, \ldots, n+1
\lab{3.1a}
\ee
We now see that thanks to the special form
 of $E$ in $A = \l E + A^{0}$, \rf{3.1} takes a simple and  equivalent form:
\be
\left(\begin{array}{ccccc}
\pa &0 &\cdots &0 &q_1 \\
0 & \pa &0 &\cdots &q_2 \\
\vdots & &\ddots & &\vdots \\ 0& & &\pa  & q_n\\
r_1& r_2 &\cdots &r_n & \pa+\l  \end{array} \right)
\left(\begin{array}{l}
{\bar \psi}_1 \\ {\bar \psi}_2 \\ \vdots \\ {\bar \psi}_n \\{\bar\psi}_{n+1}
\end{array} \right)   =        0
\lab{3.2}
\ee
The linear problem \rf{3.2} after elimination of ${\bar\psi}_{k}\;,\, k
= 1, \ldots, n$ takes a form of the scalar eigenvalue problem:
\be
- \lb \pa - \sum_{k=1}^n r_k \pa^{-1} q_k \rb {\bar\psi}_{n+1}
= \l {\bar\psi}_{n+1}
\lab{3.3}
\ee
in terms of a single eigenfunction ${\bar \psi}_{n+1}$
and the pseudo-differential operator
\be
L_n = \pa - \sum_{k=1}^n r_k \pa^{-1} q_k
\lab{3.4}
\ee
This formally relates $\GNLS_n$ hierarchy to the $SL (n+1,n)$ \cKP hierarchy
on basis of correspondence of the linear problems characterizing them.
To fully establish their complete equivalence one can show that both models
possess the same recurrence operators and therefore all the flows are
identical see \ct{affine} (\ct{cheng} describes the the case $n=1$).
We first note that the successive flows \rf{1.18} related by the
recursion operator \rf{1.18} are given by
\br
&&\pa_n \twocol{r_i}{q_l} = \cR_{(i,l),(j,m)} \twocol{r_j}{q_m} =
\lab{1.19}\\
&&\fourmat{\(-\pa + r_k \pa^{-1} q_k\)\d_{ij}+r_i\pa^{-1}q_j}
{r_i \pa^{-1} r_m +r_m\pa^{-1}r_i}{-q_l \pa^{-1} q_j -q_j\pa^{-1}q_l}
{\(\pa -q_k\pa^{-1}r_k\)\d_{lm} -q_l \pa^{-1} r_m}
\pa_{n-1} \twocol{r_j}{q_m}
\nonu
\er
{}From relation between the recursion matrix and two first bracket structure
$\cR = P_2 P_1^{-1} $ and \rf{1.19} one finds an explicit expression for
the second bracket \ct{affine} to be :
\be
P_2 = \fourmat{r_i \pa^{-1} r_j + r_j \pa^{-1} r_i}
{\(\pa -\sum_k r_k \pa^{-1} q_k \)\d_{im} - r_i \pa^{-1} q_m}
{\(\pa -\sum_k q_k \pa^{-1} r_k \)\d_{lj} - q_l \pa^{-1} r_j}
{q_l \pa^{-1} q_m + q_m \pa^{-1}q_l}
\lab{P2}
\ee
in the same basis as in \rf{1.19}.

The following Proposition proves the equivalence between
$sl (n+1)$ $\GNLS$ hierarchy defined and
the  $SL(n+1,n)$ \cKP hierarchy introduced in the previous
subsection.

\prop {\em Flows
\br
\pa_{t_m} L_n &=& \lb \( L_n^m\)_+ \, , \, L_n \rb \lab{3.15}\\
\pa_{t_m} r_i  &= & \(\(\pa - \sum_{k=1}^n r_k \pa^{-1} q_k \)^m\)_{+}r_i
 \lab{3.16}\\
\pa_{t_m} q_i  &= &- \(\(-\pa + \sum_{k=1}^n q_k \pa^{-1} r_k \)^m\)_{+} q_i
\lab{3.17}
\er
of $SL(n+1,n)$ \cKP hierarchy containing the Lax operator from \rf{3.4}
coincide with the flows produced  by the recursion operator
 $\cR$  \rf{1.19} of the $sl (n+1)$ $\GNLS$ hierarchy }

Proof is given in \ct{affine} and is accomplished by showing that both
hierarchies have identical recursion operators.
Especially \rf{3.17} yields for $m=2$:
\br
 {\pa q_i \o{\pa t_2}}& =& -\pa_x^2 q_i + 2q_i\sum_{b=1}^{n} \, q_b \, r_b
\nonumber \\
{\pa r_i\o{\pa t_2}} &=& \pa_x^2 r_i -2r_i\sum_{b=1}^{n} \,q_b \, r_b
\lab{nls}
\er
for $a=1,\ldots,n$.  These are the GNLS equations \ct{FK83},
which have been derived in \ct{affine} entirely from the AKS formalism
\ct{aks} (see also \ct{harnad}) associated with an affine Lie algebraic
structure (loop algebra $\alie \equiv \lie \otimes \IC \lb \l , \l^{-1} \rb$
with $\lie = sl(n+1)$).
This construction reveals an affine
Lie algebraic structure underlying the integrability of the
$SL (n+1, n)$ \cKP hierarchy.
\newpage
\sect{Darboux-\Back~Techniques of ${\bf SL(p,q)}$ \cKP Hierarchies and
Constrained Generalized Toda Lattices}
\subsection{Emergence of the Toda structure from Two-Bose KP System}
The two-boson KP system defined by the Lax operator
$L = D + \Phi D^{-1} \Psi \equiv D + a \( D - b\)^{-1}$
is the most basic constrained KP structure. It belongs to the $SL (2,1)$
\cKP system.

We start with the initial ``free'' Lax operator $L^{(0)} = D$ and perform a
following transformation:
\be
L^{(1)} = \(\Phi^{(0)} D {\Phi^{(0)}}^{-1}\)  \; D \;
\(\Phi^{(0)} D^{-1} {\Phi^{(0)}}^{-1}\)
\, =\, D+ \llb \Phi^{(0)} \( \ln \Phi^{(0)}\)^{\pr \pr} \rrb
D^{-1} \(\Phi^{(0)}\)^{-1}    \lab{lone}
\ee
which we call a DB transformation.
The construction below is a special application of properties
listed in Appendix B including eq. \rf{iw}.

Successive application of DB transformations leads to the following recursive
expressions:
\br
L^{(k+1)} \eq \(\Phi^{(k)}  D {\Phi^{(k)} }^{-1}\)  \; L^{(k)}
 \; \(\Phi^{(k)}  D^{-1} {\Phi^{(k)} }^{-1}\)
= D + \Phi^{(k+1)} D^{-1} \Psi^{(k+1)}
\lab{lkplus} \\
\Phi^{(k+1)} \eq \Phi^{(k)} \( \ln \Phi^{(k)}\)^{\pr \pr} + \(\Phi^{(k)}\)^2
\Psi^{(k)} \quad ,\quad \Psi^{(k+1)} = \(\Phi^{(k)}\)^{-1} \lab{pkplus}
\er
Introduce now
\be
\p_k =\ln \Phi^{(k)} \quad \to \quad  \Phi^{(k)} = e^{\p_k}  \qquad k=0,\ldots
\lab{smallp}
\ee
which allows us to rewrite \rf{pkplus} as a (ordinary one-dimensional)
Toda lattice equation:
\be
\pa^2 \p_{k}\, =\, e^{\p_{k+1} - \p_{k}} - e^{\p_{k} - \p_{k-1}}
\lab{toda}
\ee
Related objects $\psi_n$ are:
\be
\p_n = \psi_{n+1} - \psi_{n}
\lab{psieqs}
\ee
which satisfy due to eq. \rf{toda}
the following form of the Toda lattice equation:
\be
\pa^2 \psi_{n}\, =\, e^{\psi_{n+1}+\psi_{n-1}-2\psi_{n}  }
\lab{newtoda}
\ee
with $\psi_n=0$ for $n \leq 0$.
Comparing \rf{newtoda} with Jacobi's theorem \rf{jac-a} we find
the Wronskian representation for $\psi_n$:
\be
\psi_{n} = \ln  W_n \lb \P , \pa \P, \ldots , \pa^{n-1} \P \rb
\quad{\rm with}\quad \P = \P^{(0)}
\lab{newtwron}
\ee
Correspondingly $\Phi^{(k)}$ acquires the form:
\be
\Phi^{(k)} = { W_{k+1} \lb \P , \pa \P, \ldots , \pa^k \P \rb \over
W_k \lb \P , \pa \P, \ldots , \pa^{k-1} \P \rb}
\lab{phik}
\ee
We recognize at the right hand side of \rf{newtoda} a structure of
the Cartan matrix for $A_n$.
Leznov considered such an equation with Wronskian solution (in two dimensions)
in \ct{le80}.

Hence, the solutions of the (ordinary one-dimensional) Toda lattice equations,
with boundary conditions $\psi_n=0$ for $n \leq 0$, reproduce the DB
solutions of the ordinary two-boson KP hierarchy \rf{phik}
upon taking into account that $ \Phi= \Phi^{(0)} = \exp \( \p_0\) =
\exp \( \psi_1 \) $.

Note that we can write $W_n=\t_n$ with the $\tau$-function $\t_n$
satisfying Hirota's bilinear equation for the Toda lattice:
\be
\t_n \pa^2 \t_n - \( \pa \t_n \)^2 = \t_{n+1} \t_{n-1}
\lab{hirotoda}
\ee
on basis of \rf{jac-b}.

We will now find a linear discrete system (Toda spectral problem),
which leads to the above structure.
First define:
\br
R_n &=& { \Phi^{(n+1 )} \o \Phi^{(n)} } = {  \t_{n+1} \t_{n-1} \o \t_n^2 }
\lab{rndef}\\
S_n &=& \pa  \( \ln \Phi^{(n+1 )} \) = \pa  \( \ln { \t_{n+1} \o \t_n} \)
\lab{sndef}
\er
As result of \rf{hirotoda} and their definition $S_n , R_n$ satisfy
the Toda equations of motion:
\br
\pa \,S_n &= & R_{n +1} - R_n  \lab{sneq}\\
\pa \, R_n &=& R_n \( S_{n} - S_{n-1} \) \lab{rneq}
\er
These equations can also be obtained as a consistency of the spectral system
\br
\pa \Psi_n \eq \Psi_{n+1} + S_n \Psi_n     \lab{spectoda}\\
\l \Psi_n \eq \Psi_{n+1}  + S_n \Psi_n
+ R_n \Psi_{n-1}      \lab{spectodb}
\er
which defines a so-called Toda chain system.

As we will show this purely discrete system contains information
about the underlying continuous structure.
This structure is being revealed when we realize that the lattice jump
$n \to n+1$ can be given a meaning of the DB transformation.
We start by rewriting \rf{spectoda} as follows:
\be
\pa \Psi_n = \Psi_{n+1} + S_n \Psi_n  \quad \sim \quad
\Psi_{n+1}= e^{\int S_n} \pa\; e^{-\int S_n} \Psi_{n}
\lab{spectodc}
\ee
or taking into account \rf{rneq} as
\be
\Psi_{n+1}= R_n e^{\int S_{n-1}} \pa \( R_n e^{\int S_{n-1}}\)^{-1} \Psi_{n}
= \P (n) \pa \P^{-1} (n) \Psi_{n} = T (n) \Psi_{n}
\lab{todadb}
\ee
where $\P (n) =R_n e^{\int S_{n-1}} $ and $T (n)= \P (n) \pa \P^{-1} (n) $
plays a role of the DB transformation operator generating the
lattice translation $n \to n+1$.

The remaining of the Toda spectral equation \rf{spectodb}
can be given a form (with $  \Psi (n) \equiv  e^{-\int S_{n-1}} $):
\br
\l \Psi_n \eq \( \pa + R_n \( \pa - S_{n -1}\)^{-1} \)\Psi_{n} =
\llb   \pa + R_n e^{\int S_{n-1}} \pa^{-1}  e^{-\int S_{n-1}} \rrb  \Psi_{n}
\nonu \\
\eq \llb   \pa + \P (n) \pa^{-1}  \Psi (n) \rrb  \Psi_{n}
= L (n)  \Psi_{n}
\lab{specln}
\er
of the Lax eigenvalue problem of the two-bose KP Lax system with
generic two-bose KP Lax operator $ L = D + \P D^{-1}  \Psi $.
Hence the Toda lattice spectral problem has been shown equivalent
to the continuous \cKP system possessing the symmetry with respect to
the DB transformations.

\subsection{On the DB Transformations of the ${\bf SL(r+n,n)}$ \cKP
Lax Operators}

We shall here consider behavior of the general class of constrained KP
Lax operators from $SL(r+n,n)$ \cKP hierarchy under
an arbitrary DB transformation $ {\wti L} =\chi D \chi^{-1} L
\chi D^{-1} \chi^{-1} $ where $\chi$ is an eigenfunction of the Lax operator
$L$ as given in \rf{f-5}.
The transformed Lax operator reads:
\br
{\wti L} &=& \chi D \chi^{-1} \( L_{+} + \sum_{i=1}^{n} \Phi_i D^{-1} \Psi_i \)
\chi D^{-1} \chi^{-1} \equiv {\wti L}_{+} + {\wti L}_{-}  \lab{baker-1} \\
{\wti L}_{+} &=& {L}_{+} + \chi \( \pa_x \( \chi^{-1} L_{+} \chi \)_{\geq 1}
D^{-1}\) \chi^{-1}    \lab{baker-2}  \\
{\wti L}_{-} &=& {\wti \Phi}_0 D^{-1} {\wti \Psi}_0 +
\sum_{i=1}^{n} {\wti \Phi}_i D^{-1} {\wti \Psi}_i   \lab{baker-3} \\
{\wti \Phi}_0 &=& \chi \llb \pa_x \( \chi^{-1} L_{+} \chi \) + \sum_{i=1}^n
\( \pa_x \( \chi^{-1}\Phi_i\) \pa_x^{-1} \( \Psi_i \chi\)
+ \Phi_i \Psi_i \) \rrb \equiv \( \chi D \chi^{-1} L \) \chi  \lab{baker-4} \\
{\wti \Psi}_0 &=& \chi^{-1} \qquad ,\quad
{\wti \Phi}_i = \chi \pa_x \( \chi^{-1} \Phi_i \) \qquad ,\quad
{\wti \Psi}_i = - \chi^{-1} \pa_x^{-1} \( \Psi_i \chi \)   \lab{baker-5}
\er
{}From the above discussion we know that all involved functions are
(adjoint) eigenfunctions of $L$ \rf{f-5}, {\sl i.e.}, they satisfy:
\be
\partder{}{t_k} f = L^{{k\over r}}_{+} f
\qquad f = \chi ,\Phi_i \quad ; \quad
\partder{}{t_k} \Psi_i = - {L^\ast}^{{k\over r}}_{+} \Psi_i
\lab{baker-6}
\ee
We are interested in the special case when $\chi$ coincides with one of the
original eigenfunctions of $L$, {\sl e.g.}  $\chi = \Phi_1$. Then
${\wti \Phi}_1 =0$ and the DB transformation \rf{baker-1} preserves the
form \rf{f-5} of the Lax operators involved,
{\sl i.e.}, it becomes an {\em auto}-\Back ~transformation.
Applying the successive DB transformations in this case yields:
\br
L^{(k)} \eq T^{(k-1)} L^{(k-1)} \( T^{(k-1)}\)^{-1} =
\( L^{(k)}\)_{+} + \sum_{i=1}^n \Phi_i^{(k)} D^{-1} \Psi_i^{(k)} \;\;, \;\;
T^{(k)} \equiv \Phi_1^{(k)} D \( \Phi_1^{(k)}\)^{-1}  \phantom{aaaa}
\lab{shabes-3} \\
\Phi_1^{(k+1)} \eq \( T^{(k)} L^{(k)}\) \Phi_1^{(k)} \quad ,\quad
\Psi_1^{(k+1)} = \( \Phi_1^{(k)}\)^{-1}
\qquad, \qquad k =0,1,\ldots \lab{shabes-2} \\
\Phi_i^{(k+1)} \eq T^{(k)} \Phi_i^{(k)} \equiv
\Phi_1^{(k)} \pa_x \( \( \Phi_1^{(k)}\)^{-1} \Phi_i^{(k)}\)  \lab{shabes-5} \\
\Psi_i^{(k+1)} \eq - \( \Phi_1^{(k)}\)^{-1} \pa_x^{-1}
\( \Psi_i^{(k)}\Phi_1^{(k)} \) \qquad , \;\; i=2,\ldots , n  \lab{shabes-5a}
\er
Using the first identity from \rf{shabes-3}, {\sl i.e.}, $ L^{(k+1)} T^{(k)}=
T^{(k)} L^{(k)}$ , one can rewrite \rf{shabes-2} in the form:
\be
\Phi_1^{(k)} = T^{(k-1)} T^{(k-2)} \cdots T^{(0)}
\( \( L^{(0)}\)^k \Phi_1^{(0)}\)
\lab{shabes-4}
\ee
whereas:
\be
\Phi_i^{(k)} = T^{(k-1)} T^{(k-2)} \cdots T^{(0)} \Phi_i^{(0)} \qquad ,\;\;
\quad i=2,\ldots , n  \lab{shabes-6}
\ee
Finally, for the coefficient of the next-to-leading differential term in
\rf{f-5} ~$ u_{r-2} = r \, Res\, L^{{1\over r}} = r\, \pa_x^2 \ln \t$, we
easily obtain from \rf{baker-2} (with $\chi =\Phi_1$) its $k$-step
DB-transformed expression:
\be
{1\over r} \( u_{r-2}^{(k)} - u_{r-2}^{(0)}\) =
\pa_x^2 \ln \frac{\t^{(k)}}{\t^{(0)}} =
\pa_x^2 \ln \( \Phi_1^{(k-1)} \cdots \Phi_1^{(0)}\)   \lab{sol-3-a}
\ee

\mark
Let us particularly stress that the eigenfunctions we are working with
{\em are not} Baker-Akhiezer eigenfunctions of $L$ from \rf{f-5}.
Imagine, namely that we started with $L = L_{+} + \psi D^{-1} \psi^{\ast}$,
where $\psi, \psi^{\ast}$ are BA functions.
Choosing $\chi=\psi$ would result according to \rf{baker-4} and \rf{baker-5}
in ${\wti L} = D$. Hence in such case we would be able
to transform away the pseudo-differential part of the \cKP Lax operator
by a finite number of DB transformations.

\subsection{DB Transformation and Eigenfunctions of the Lax Operator}
After seeing DB transformation in action in the simple cases shown in the first
two subsections of this chapter we are now ready to review the basic properties
of DB transformation from the point of view of preserving the form of
the Lax evolution equation. Related material can be found in e.g.
\ct{chau,oevela}.

\lemma {\em For arbitrary pseudo-differential operator $A$ we have the
following identity \ct{oevelr}:}
\be
\( \chi D \chi^{-1} A \chi D^{-1} \chi^{-1} \)_{+} =
\chi D \chi^{-1} \( A \)_{+}  \chi D^{-1} \chi^{-1}
-  \chi \pa_x \(\chi^{-1} (( A )_{+} \chi) \)  D^{-1} \chi^{-1}
\lab{aonchi}
\ee
For $L$ satisfying a general KP-KdV Lax equation $\pa_k  L =
\sbr{L_{+}^{k\o r}}{L}$ the transformed Lax operator
${\ti L} \equiv T L T^{-1}$ will satisfy:
\be
\pa_k {\ti L} =
\sbr{T L_{+}^{k\o r}T^{-1} + (\pa_k T )T^{-1}}{{\ti L}}
\lab{tisato}
\ee
Let $\Phi$ be an eigenfunction of $L$ i.e. $L_{+}^{k\o r} \Phi =
\pa_k \Phi$, which enters the DB transformation
through $T = \Phi D \Phi^{-1}$.
One verifies easily that for $A= L^{k\o r}$ and $\chi = \Phi$
equation \rf{aonchi} becomes
\be
\( T L^{k\o r}T^{-1} \)_{+} =     T L_{+}^{k\o r}T^{-1} + (\pa_k T )T^{-1}
\lab{lonphi}
\ee
Correspondingly \rf{tisato} takes the form:
\be
\pa_k {\ti L} =
\sbr{{\ti L}_{+}^{k\o r}}{{\ti L}}
\lab{tisato-a}
\ee
and we have therefore established:

\prop {\em The DB transformation with an eigenfunction $\Phi$
preserves the form of the Lax equation \rf{lax-eq} i.e.
the DB transformed Lax operator satisfies the same evolution
equation as the original Lax operator}.

For the pseudo-differential operator $A= D + a_1 D ^{-1} + \ldots$ we have
\be
{\rm Res} \( \chi D \chi^{-1} A \chi D^{-1} \chi^{-1} \)=
 \( \ln \chi \)^{\pr \pr} + {\rm Res} \( A \)
\lab{resa}
\ee
For the case of the Lax operator $L$ with the $\t$-function satisfying
${\rm Res} L = \pa^2 \ln \t$ we get therefeore \ct{chau}:

\prop {\em Under the DB transformation with an eigenfunction $\Phi$
the $\t$-function associated to the Lax operator $L$
transforms according to $\t \to {\wti \t} = \P \t$}.

\subsection{Exact Solutions of ${\bf SL(p,q)}$ \cKP hierarchy via DB
Transformations}
Armed with the Wronskian identities from Appendix B, we can represent the
$k$-step DB transformation \rf{shabes-4}---\rf{sol-3-a} in terms of
Wronskian determinants involving the coefficient functions of the ``initial''
Lax operator
\be
L^{(0)} = D^r + \sum_{l=0}^{r-2} u_l^{(0)} D^l +
\sum_{i=1}^n \Phi_i^{(0)} D^{-1} \Psi_i^{(0)}    \lab{seq-b}
\ee
only.
Indeed, using identity \rf{iw} and defining:
\be
\(L^{(0)}\)^k \P_1^{(0)} \equiv \chi^{(k)} \qquad \; k=1,2,\ldots
\lab{defchi}
\ee
we arrive at the following general result:

\prop
{\em The $k$-step DB-transformed eigenfunctions and the tau-function
\rf{shabes-4}---\rf{sol-3-a} of the $SL(r+n,n)$ \cKP system \rf{iss-8aa} for
arbitrary initial $L^{(0)}$ \rf{seq-b} are given by:
\br
\P_1^{(k)}&=& \frac{W_{k+1}\lb \P_1^{(0)},\chi^{(1)},\ldots, \chi^{(k)}\rb}
{W_{k}\lb \P_1^{(0)}, \chi^{(1)},\ldots ,\chi^{(k-1)}\rb}  \lab{pchi-a}  \\
\P_j^{(k)}&=&\frac{W_{k+1} \lb \P_1^{(0)},\chi^{(1)},\ldots ,\chi^{(k-1)},
\P_j^{(0)}\rb}{W_{k}\lb \P_1^{(0)}, \chi^{(1)}, \ldots,  \chi^{(k-1)}\rb}
\qquad , \;\; j=2,\ldots ,n    \lab{pchi-aa}  \\
\tau^{(k)} &=& W_{k} \lb  \P_1^{(0)}, \chi^{(1)},
\ldots,  \chi^{(k-1)}\rb \tau^{(0)}      \lab{tauok}
\er
where $\tau^{(0)}, \tau^{(k)}$ are the $\tau$-functions of $L^{(0)},
L^{(k)}$ , respectively, and $\chi^{(i)}$ is given by \rf{defchi}.}
\mskp

\underbar{Example} Consider $SL(2,1)$ \cKP hierarchy with the Lax operator:
$L^{(0)} = D + \Phi^{(0)} D^{-1} \Psi^{(0)}$, which serves as a starting
point of the successive DB transformations:
\br
L^{(k)} &=& D + \Phi^{(k)} D^{-1} \Psi^{(k)}=
T^{(k-1)} L^{(k-1)} \( T^{(k-1)}\)^{-1} \lab{osol-1} \\
\Phi^{(k)} & = & T^{(k-1)} L^{(k-1)} \Phi^{(k-1)}  \lab{osol-2}
\er
By iteration we find:
\br
\Phi^{(k)} \eq T^{(k-1)} \cdots T^{(1)} T^{(0)} \( (L^{(0)})^{k} \Phi^{(0)} \)
=\frac{W_{k+1}\lb \P^{(0)},\chi^{(1)},\ldots, \chi^{(k)}\rb}
{W_{k}\lb \P^{(0)}, \chi^{(1)},\ldots ,\chi^{(k-1)}\rb} \lab{osol-3} \\
\Psi^{(k)} \eq \(\Phi^{(k-1)}\)^{-1}   \lab{osol-4}
\er
Comparing two alternative expressions involving the tau-function
$\t^{(k)}$:
\be
\t^{(k)} = \P^{(k)} \cdots \P^{(1)} \P^{(0)} \t^{(0)} =
W_{k+1}\lb \P_1^{(0)},\chi^{(1)},\ldots, \chi^{(k)}\rb \t^{(0)}
\lab{tkt0}
\ee
and ${\rm Res} L^{(k)} = \Phi^{(k)}   \Psi^{(k)}=  \pa^2 \ln \t^{(k)}$
we arrive at:
\br
\Phi^{(0)}   \Psi^{(0)}&=&-
\pa^2 \ln W_{k}\lb \P^{(0)}, \chi^{(1)},\ldots ,\chi^{(k-1)}\rb
\lab{genhiro}\\
&+&  \frac{W_{k+1}\lb \P^{(0)},\chi^{(1)},\ldots, \chi^{(k)}\rb
W_{k-1}\lb \P^{(0)},\chi^{(1)},\ldots, \chi^{(k-2)}\rb}
{W_{k}^2\lb \P^{(0)}, \chi^{(1)},\ldots ,\chi^{(k-1)}\rb} \nonu
\er
which generalizes the structure of the Hirota equation \rf{hirotoda}
found for the case $ \Phi^{(0)} =  \Psi^{(0)}=0$ in subsection
3.1.

\underbar{Example:} Construction of the $SL(3,1)$ \cKP Lax operator, {\sl
i.e.},
$r=2, n=1$ in \rf{f-5} i.e.
\be
L = D^2 + u + A \( D - B\)^{-1} = D^2 + u + \Phi D^{-1} \Psi \lab{sl-31}
\ee
starting from the ``free'' $L^{(0)}= D^2$.
This example is pertinent to the simplest nontrivial string
two-matrix model \ct{enjoy}.

{}From the basic formulas for successive DB transformations
\rf{shabes-2}--\rf{shabes-3}, applied to \rf{sl-31}, we have:
\br
L^{(k)} \eq D^2 + u^{(k)} + \Phi^{(k)} D^{-1} \Psi^{(k)}  \lab{sol-1} \\
L^{(k)} \!\!&\to&\!\!  L^{(k+1)} = T^{(k)} L^{(k)} \( T^{(k)}\)^{-1}
\quad ,\quad  T^{(k)} = \Phi^{(k)} D \( \Phi^{(k)}\)^{-1} \lab{sol-2} \\
u^{(k)} \eq 2 Res\, L^{\h} \equiv 2 \pa_x^2 \ln \t^{(k)} =
2 \pa_x^2 \ln \( \Phi^{(k-1)} \cdots \Phi^{(0)}\)  \lab{sol-3} \\
\Phi^{(k)} \!\!\!&\equiv& \!\!\!A^{(k)} e^{ \int B^{(k)}} =
\Phi^{(k-1)} \llb \pa_x \( {1\over {\Phi^{(k-1)}}} \pa_x^2
\Phi^{(k-1)} + 2 \pa_x^2 \ln \( \Phi^{(k-2)} \cdots \Phi^{(0)}\) \) +
\frac{\Phi^{(k-1)}}{\Phi^{(k-2)}} \rrb \phantom{aaa} \lab{sol-4} \\
\Psi^{(k)} \!\!&\equiv& \!\!
e^{ -\int B^{(k)}} = \( \Phi^{(k-1)}\)^{-1} \lab{sol-5} \\
u^{(0)} \eq 0 \quad ,\quad \Psi^{(0)} = 0 \quad ;\quad
\Phi^{(0)} = \int_{\Gamma} \frac{d\l}{2\pi} c(\l ) e^{\xi \( \l ,\{ t\}\)}
\quad ,\;\;
\xi \( \l ,\{ t\}\) \equiv \l x + \sum_{j \geq 2} \l^{j} t_j \lab{sol-6}
\er
where $\Phi^{(0)}$ in \rf{sol-6} is an arbitrary eigenfunction of the ``free''
$L^{(0)}= D^2$ (the contour $\Gamma$ in the complex $\l$-plane is chosen such
that the generalized Laplace transform of $c(\l )$ is well-defined).

As a corollary from the above proposition, we get in the case of \rf{sol-1} :
\br
\Phi^{(k)}& =& T^{(k-1)} \cdots T^{(0)} \( \pa_x^{2k} \Phi^{(0)} \) =
\frac{W \llb \Phi^{(0)}, \pa_x^2 \Phi^{(0)},
\ldots , \pa_x^{2k}\Phi^{(0)}\rrb}{ W \llb \Phi^{(0)}, \pa_x^2 \Phi^{(0)},
\ldots , \pa_x^{2(k-1)}\Phi^{(0)} \rrb} \lab{sol-7a} \\
\t^{(k)}& =& W \llb \Phi^{(0)}, \pa_x^2 \Phi^{(0)},
\ldots , \pa_x^{2(k-1)}\Phi^{(0)} \rrb  \lab{sol-8}
\er

Substituting \rf{sol-7a},\rf{sol-8} into \rf{sol-3}--\rf{sol-5} we obtain
the following explicit solutions for the coefficient functions of \rf{sl-31} :
\br
u^{(n)} &= &2 \pa_x^2 \ln W \llb \Phi^{(0)}, \pa_x^2 \Phi^{(0)},
\ldots , \pa_x^{2(n-1)}\Phi^{(0)} \rrb \phantom{aa} \lab{sol-3-aa} \\
B^{(n)} &=& \pa_x \ln \( \frac{W \llb \Phi^{(0)},
\pa_x^2 \Phi^{(0)},\ldots , \pa_x^{2(n-1)}\Phi^{(0)} \rrb}{W \llb
\Phi^{(0)}, \pa_x^2 \Phi^{(0)},\ldots , \pa_x^{2(n-2)}\Phi^{(0)} \rrb}\)
\lab{sol-3-b} \\
A^{(n)}&=&\frac{W \llb \Phi^{(0)},
\pa_x^2 \Phi^{(0)},\ldots , \pa_x^{2n}\Phi^{(0)} \rrb \, W \llb
\Phi^{(0)}, \pa_x^2 \Phi^{(0)},\ldots , \pa_x^{2(n-2)}\Phi^{(0)} \rrb}{\( W
\llb \Phi^{(0)},\pa_x^2 \Phi^{(0)},\ldots , \pa_x^{2(n-1)}\Phi^{(0)} \rrb\)^2}
\lab{sol-4-b}
\er

In the more general case of
$SL(r+1,1)$ \cKP Lax operator for arbitrary finite $r$ :
\be
L = D^r + \sum_{l=0}^{r-2} u_l D^l + \Phi D^{-1} \Psi  \lab{sol-9}
\ee
which defines the integrable hierarchy corresponding to the general string
two-matrix model (cf. \ct{enjoy,office}), the generalizations of \rf{sol-7a}
and \rf{sol-8} read:
\br
\Phi^{(k)} \eq T^{(k-1)} \cdots T^{(0)} \( \pa_x^{k\cdot r} \Phi^{(0)} \)
= \frac{W \llb \Phi^{(0)}, \pa_x^{r} \Phi^{(0)},
\ldots , \pa_x^{k\cdot r}\Phi^{(0)} \rrb}{W \llb \Phi^{(0)},
\pa_x^{r} \Phi^{(0)}, \ldots , \pa_x^{(k-1)\cdot r}\Phi^{(0)} \rrb}
\lab{sol-10} \\
{1\over r} u_{r-2}^{(k)} \eq Res\, L^{{1\over r}} = \pa_x^2 \ln \t^{(k)} \quad
,
\quad  \t^{(k)} = W \llb \Phi^{(0)}, \pa_x^{r} \Phi^{(0)},
\ldots , \pa_x^{(k-1)\cdot r}\Phi^{(0)} \rrb  \lab{sol-11}
\er
where $\Phi^{(0)}$ is again given explicitly by \rf{sol-6}.

\subsection{Relation to the Constrained Generalized Toda Lattices}
Here we shall establish the equivalence between the set of successive DB
transformations of the $SL(r+1,1)$ \cKP system \rf{sol-9} :
\br
L^{(k+1)} &=&  T^{(k)}\; L^{(k)} \; \(T^{(k)}\)^{-1} \qquad ,\quad
T^{(k)} = \Phi^{(k)}  D {\Phi^{(k)} }^{-1}       \lab{seq-a} \\
L^{(0)} &=& D^r + \sum_{l=0}^{r-2} u_l^{(0)} D^l +
\Phi^{(0)} D^{-1} \Psi^{(0)}    \lab{seq-bb}
\er
and the equations of motion of a {\em constrained}
generalized Toda lattice system, underlying the two-matrix string model,
which contains, in particular, the {\em two-dimensional} Toda lattice
equations.

For simplicity we shall illustrate the above property
on the simplest nontrivial case of $SL(3,1)$ \cKP hierarchy
\rf{sl-31}.
We note that eqs.\rf{sol-3}--\rf{sol-5} (or \rf{sol-3-aa}--\rf{sol-4-b}) can
be cast in the following recurrence form:
\br
\pa_x \ln A^{(n-1)}& =& B^{(n)} - B^{(n-1)}  \lab{Tod-1} \\
u^{(n)} - u^{(n-1)}& =& 2\pa_x B^{(n)}  \lab{Tod-2} \\
A^{(n)} - A^{(n-1)} &=& \pa_x \( \( B^{(n)}\)^2 + \h \( u^{(n)} + u^{(n-1)}\)\)
\lab{Tod-3}
\er
with ``initial'' conditions (cf. \rf{sol-6}) :
\be
A^{(0)}=B^{(0)}=u^{(0)}=0 \qquad , \quad B^{(1)} \equiv \pa_x \ln \Phi
\lab{Tod-4}
\ee
where $\Phi$ is so far an arbitrary function. Now, we can view
\rf{Tod-1}--\rf{Tod-3} as a system of lattice equations for the dynamical
variables $A^{(n)},B^{(n)},u^{(n)}$ associated with each lattice site $n$
and subject to the boundary conditions:
\be
A^{(n)}=B^{(n)}=u^{(n)}=0 \quad ,\;\; n \leq 0  \lab{bound-cond}
\ee
Taking \rf{Tod-4} as initial data, one can solve the lattice system
\rf{Tod-1}--\rf{Tod-3} step by step (for $n=1,2,\ldots$) and the
solution has precisely the form of \rf{sol-3-aa}--\rf{sol-4-b}.

\subsection{Darboux-B\"{a}cklund Transformation and the Dressing Chain.}
As a small digression we will here apply Darboux-B\"{a}cklund (DB)
transformation to the KdV hierarchy
emphasizing similarity with the approach developed in the KP setting.
Recall from \rf{sol-1}-\rf{sol-6} that for $T = \Phi D \Phi^{-1}$ we have
\be
T \, D^2 T^{-1} = D^2 + 2 \pa^2_x \ln \Phi  + \Phi \( \Phi^{-1} \Phi^{\pr
\pr}\)^{\pr} D^{-1} \Phi^{-1}
\lab{dtwo}
\ee
The inverse scattering problem corresponding to the KdV equation
$\pa_t u + 3 u \pa u +\h \pa^3 u=0$ is defined in terms of the
differential operator $L= D^2 + u$, which transforms as:
\br
L_1 \equiv T \, \( D^2 +u\) T^{-1} \eq D^2 + 2 \pa^2_x \ln \Phi  +
\Phi \( \Phi^{-1} \Phi^{\pr \pr}\)^{\pr} D^{-1} \Phi^{-1} \nonu \\
&+& u + \Phi u^{\pr} D^{-1} \Phi^{-1}
\lab{dtwou}
\er
In order for $L_1$ to be a differential operator we have to demand that
the pseudo-differential part in \rf{dtwou} vanishes:
\be
\Phi \( \Phi^{-1} \Phi^{\pr \pr}\)^{\pr} + \Phi u^{\pr} =0
\lab{demand}
\ee
which translates into relation between $u$ and $\P$:
\be
u = - { \Phi^{\pr \pr} \o \Phi} - \l_0 = - f_0^2 - f_0^{\pr} - \l_0
\lab{deman-a}
\ee
where $\l_0$ is an integration constant and $ f_0 \equiv ( \ln \Phi )^{\pr}$.
In this notation:
\be
L = D^2 +u = D^2 - f_0^2 - f_0^{\pr} - \l_0 = \( D + f_0\)  \( D - f_0\)
- \l_0
\lab{dtwof}
\ee
Since $ T = D - f_0$ and $ T^{\dag} = -D - f_0$
we can factorize the Lax operator $L$ according to
\be
L =- T^{\dag} \, T + \l_0 \lab{factorl}
\ee
In this notation the DB transformation on $L$:
\be
L_1 = T  L T^{-1}  = - T \, T^{\dag} - \l_0 =
D^2 - f_0^2 +f_0^{\pr} - \l_0  = D^2 + u_1
\lab{kdvdb}
\ee
amounts to reversing the
$T,T^{\dag}$ operators in expression for $L$. Moreover we find that
\be
u_1 = - f_0^2 +f_0^{\pr} - \l_0  = - f_1^2 -f_1^{\pr} - \l_1
    = u + 2 ( \ln \Phi)^{\pr\pr}
\lab{uone}
\ee
where we introduced new variables $f_1 = ( \ln \Phi_1 )^{\pr}$ and
$\l_1$, which allow us to rewrite $L_1$ as $L_1 =- T^{\dag}_1 \, T_1 + \l_1$
with $ T_1 = D - f_1$.
Clearly we have $L_1 \Phi_1 = - \l_1 \Phi_1$.

Successive use of the DB transformations \ct{dress-s} leads to the Lax
operators
$L_n = D^2 +u_n$ with
\be
u_n = - f_{n-1}^2 +f_{n-1}^{\pr} - \l_{n-1}  = - f_n^2 -f_n^{\pr} - \l_n
    = u + 2 ( \ln \Phi\cdots \Phi_{n-1})^{\pr\pr}
\lab{unn}
\ee
with a string of eigenvalue problems:
\be
L_i \P_i = - \l_i \Phi_{i }\quad {\rm or} \quad
\( \pa^2 + u_i + \l_i\) \Phi_{i } = 0 \qquad
\Phi\equiv \Phi_0 \quad, \quad u_0 \equiv u
\lab{sturm}
\ee
for $i=0,1,\ldots$.
Let $\Psi_1, \ldots, \Psi_n$ be solutions of the related eigenvalue problem:
\be
L_0 \Psi_i = - \l_i \Psi_{i }\quad {\rm or} \quad
\( \pa^2 + u + \l_i\) \Psi_{i } = 0 \qquad
\Psi_0 \equiv \Phi  \;, \quad i=0,1,\ldots
\lab{sturm-a}
\ee
We have a following theorem shown by Crum \ct{crum}, which establishes
relation between $\P_i$ and $\Psi_i$ eigenfunctions.

\prop {\em The function}
\be
\Phi_{n} \equiv { W_{n+1} \lb \Psi_0, \Psi_1, \ldots, \Psi_{n}\rb \o
W_{n } \lb \Psi_0, \Psi_1, \ldots, \Psi_{n-1}\rb }
\lab{crum-a}
\ee
{\em satisfies the differential equation}:
\br
&&\( \pa^2 + u_n + \l_n\) \Phi_{n} = 0 \lab{crum-b}\\
&&u_n = u+ 2 \pa^2 \ln W_{n } \lb \Psi_0, \Psi_1, \ldots, \Psi_{n-1}\rb
\lab{crum-c}
\er
\proof To prove it let us notice that $ L_n = (T_{n-1} \cdots T_0) L_0
(T_{n-1} \cdots T_0)^{-1}$, with $T_i = \Psi_i \pa \Psi_i^{-1}$.
{}From $L_n \Phi_n = -\l_n \Phi_n $
we find that $\Phi_n = T_{n-1} \cdots T_0 \Psi_n$.
By use of induction and \rf{iw} one completes the proof.
Equation \rf{crum-c} follows automatically from \rf{crum-a} and \rf{unn}.

Note a clear analogy of the dressing chain construction with the Susy QM
\ct{matveev}.
Define namely, $T_n = D-f_n$ with $T_n^{\dag} = -(D+f_n)$ and introduce
$H_{\pm}$ through:
\be
T_n T_n^{\dag}= - (D^2 - f_n^2 +f_n^{\pr} ) = H_{-} - \l_n
\quad ;\quad T_n^{\dag} T_n = - (D^2 - f_n^2 -f_n^{\pr} ) = H_{+} - \l_n
\lab{susy-a}
\ee
These relations can be cast into the QM Susy algebra:
\be
\pbr{Q_{-}}{Q_{+}} = H- \l_n \qquad ;\qquad \sbr{Q_{\pm}}{H} = 0
\lab{susy-b}
\ee
in terms of the $2 \times 2$ matrices:
\be
Q_{+} = \fourmat{0}{0}{T}{0} \quad ; \quad Q_{-} = \fourmat{0}{T^{\dag}}{0}{0}
\quad ; \quad
H =\fourmat{H_{-}}{0}{0}{H_{+}}
\lab{susy-c}
\ee
The eigenvalue problem $H V_n = \l_n V_n$ with $V_n = ( V_{-}, V_{+} )^{T}$
takes a familiar form $(D^2 - f_n^2 I + f_n^{\pr} \sigma_3 ) V= 0$
and solutions are given by
\be
V_{\pm} = \exp{ \( \pm  \int f_n\) }= ( \Phi_n)^{\pm 1}
\lab{susy-d}
\ee
as verified by rewriting the eigenvalue problem as
$ T^{\dag}T\, V_{+}=0$ and $ T T^{\dag}\, V_{-}=0$.
Hence the Darboux techniques prove useful in constructing exact solution of
the Schr\"{o}dinger problems of the supersymmetric quantum mechanics.

\subsection{Connection to Grassmannian Manifolds and
n-Soliton Solution for the KP Hierarchy}.

Let $\{ \psi_1, \ldots, \psi_n \}$ be a basis of solutions of the $n$-th
order equation $L \psi = 0$,
where $L=\( D + v_n \) \( D + v_{n-1} \) \ldots \( D + v_1 \) $.
If $W_k$ denotes the Wronskian determinant of $\{ \psi_1, \ldots, \psi_k\}$
one can then show that \ct{wilson85,ince}:
\be
v_i = \pa \( \ln { W_{i-1} \over W_{i}} \) \qquad W_0 =1
\lab{wil-a}
\ee
This allows to establish that the space of differential operators
is parametrized by the Grassmannian manifold (see e.g. \ct{wilson85,matsuo}).
Start namely with the given differential operator
$L_n= D^n + u_1 D^{n-1} + \cdots + u_n$ and
determine the kernel of $L_n$ given by $n$-dimensional subspace
of some Hilbert space of functions $\cH$, spanned, let say, by
$\{ \psi_1, \ldots, \psi_n \}$. This establishes the connection one way.
On the other hand let $\{ \psi_1, \ldots, \psi_n \}$ be a basis of one point
$\O$ of a Grassmannian manifold ${\rm Gr}^{(n)}$.
Define the differential equation as $L_n (\O) f =  W_{k} (f) / W_k$.
{}From \rf{kw} this associates the differential operator
\be
L_n= D^n + u_1 D^{n-1} + \cdots + u_n=\( D + v_n \) \( D + v_{n-1} \)
\cdots \( D + v_1 \)
\lab{miura}
\ee
given by a Miura correspondence to a given point of the Grassmannian.

Let us now comment on connection to n-soliton solution for the KP hierarchy.
Assume that the above functions $\psi_i\; i =1 , \ldots, n$ have the
property $\pa_m \psi_i = \pa^m \psi_i $ for arbitrary $m \geq 1$
{}~($\pa_m \equiv \partder{}{t_m}$),
in other words $\psi_i$ are eigenfunctions of $L^{(0)}=D$.
We introduce $ L \equiv L_n D L_n^{-1}$, where $L_n$ is defined in terms
of $\{ \psi_1, \ldots, \psi_n \}$ as in \rf{wil-a} and \rf{miura}.
It is known \ct{Zakh,dickey-b,ohta}
that such a Lax operator satisfies a generalized Lax equation
$\pa_m L = \lb L^m_{+}, L \rb$.

Using \rf{iw} we can rewrite the above Lax operator as a result of successive
DB transformations applied on $D$:
\be
L = L_n D L_n^{-1} = T_n \, T_{n-1}\, \cdots\, T_1 \, D
T_1^{-1} \cdots\, T_{n-1}^{-1} \, T_n^{-1}
\lab{nsolax}
\ee
where $T_i$ are given in terms of Wronskians as in \rf{transf}.
It follows that $L$ can be cast in a form of the Lax operator
belonging to subclass of the  $SL (1+n,n)$ \cKP hierarchy and
having the form as in \rf{f-5} with $r=1$.
Using the formalism developed in this paper one can prove by induction that
the corresponding $\tau$-function of $L$ takes a Wronskian form
$ \tau_n = W_n \lb \psi_1, \ldots, \psi_n \rb$ reproducing n-soliton solution
to the KP equation derived in \ct{nimmo}.
In fact, choosing $\psi_i = \exp \( \sum t_k \a^k_i\) +
\exp \( \sum t_k \b^k_i\)$
allows to rewrite $\tau_n $ in the conventional form of the
$n$-soliton solution to the KP equation \ct{hirota,fritz}.
\newpage
\sect{Two-Matrix Model as a ${\bf SL (r+1,1)}$-\cKP Hierarchy}
\subsection{Two-Matrix model, Orthogonal Polynomials Technique}
We shall consider the two-matrix model with partition function :
\be
Z_N \lb t,{\ti t},g \rb = \int dM_1 dM_2 \exp \lcurl
\sum_{r=1}^{p_1} t_r \Tr M_1^r +
\sum_{s=1}^{p_2} {\ti t}_s \Tr M_2^s + g \Tr M_1 M_2 \rcurl   \lab{2-1}
\ee
where $M_{1,2}$ are Hermitian $N \times N$ matrices,
and the orders of the matrix ``potentials'' $p_{1,2}$ may be
finite or infinite.
As in ref.\ct{BX9212} and \ct{shaw,ahns} we will use the method
of generalized orthogonal polynomials \ct{ortho-poly} to evaluate partition
function \rf{2-1}.
After angular integration in \rf{2-1} we obtain
\be
Z_N \lb t,{\ti t},g \rb ={ 1 \o N !} \int \prod_{i=1}^N d \l_i d {\ti \l}_i
 \exp \lcurl \sum_{i=1}^{N}\( V (\l_i) + {\ti V} ({\ti \l}_i) +
 g \l_i {\ti \l}_i\) \rcurl \Delta (\l_i) \Delta ({\ti \l}_i)
\lab{partl}
\ee
where $ V ( \l) = \sum t_k \l^k $, ${\ti V} ({\ti \l}) = \sum {\ti t}_s
{\ti \l}^s$.
$\Delta (\l_i), \Delta ({\ti \l}_i)$ are standard Van der
Monde determinants. As in one-matrix model one can deal with them
using orthogonal polynomials.
Here we will work with two families of the orthogonal polynomials:
\be
P_n (\l_1) = \l_1^n + O \(\l_1^{n-1}\)\quad;\quad
{\ti P}_m (\l_2) = \l_2^m+ O \(\l_2^{m-1}\)\quad,\quad
n,m= 0,1,\ldots
\lab{opoly}
\ee
which enter the orthogonal relation:
\be
\int d \l_1 d \l_2  \exp \( V (\l_1) + {\ti V} (\l_2) +
 g \l_1 \l_2 \) P_n (\l_1) {\ti P}_m (\l_2) = h_m \d_{nm}
\lab{orthorel}
\ee
Following approach to one-matrix model we write down the recursion relations
for the polynomials in \rf{opoly}.
\be
\l_1 P_n (\l_1) = \sum_{l=0}^{n+1} Q_{nl} P_l (\l_1) \qquad  ; \qquad
\l_2 {\ti P}_n (\l_2 ) = \sum_{l=0}^{n+1}{\ti P}_l (\l_2 ) {\wti Q}_{ln}
\lab{recurpo}
\ee
{}From the definition of orthogonal polynomials it follows that
\be
Q_{n,n+1} =1 \; ; \; Q_{n,l} = 0 \;\;\;\; l \geq n+2 \quad \mbox{and} \quad
{\wti Q}_{n+1,n} =1 \; ; \; {\wti Q}_{l,n} = 0 \;\;\; \; l \geq n+2
\lab{qtiqone}
\ee
The orthogonal polynomials approach leads to expression for the partition
function in terms of $h_n$:
\be
Z_N = h_0 h_1 \cdots h_{N-2} h_{N-1}
\lab{partfctz}
\ee
{}From the orthogonal relation \rf{orthorel} and definitions \rf{qtiqone}
we obtain:
\be
\partder{\ln h_n}{t_r} = \(Q^r\)_{nn} \qquad ; \qquad
\partder{\ln h_n}{{\ti t}_s} = \({\wti Q}^s\)_{nn}
\lab{hnttit}
\ee
as well as
\br
\partder{P_n}{t_r} &=& - \sum_{l=0}^{n-1} Q^r_{nl} P_l (\l_1)
\quad ; \quad
\partder{{\ti P}_n}{t_r} = - \sum_{l=0}^{n-1} {\ti P}_l (\l_2) (H^{-1} Q^r
H)_{ln}   \lab{pntr}\\
\partder{P_n}{{\ti t}_s} &=& - \sum_{l=0}^{n-1} {\bar Q}^s_{nl} P_l (\l_1)
\quad ; \quad
\partder{{\ti P}_n}{{\ti t}_s} = - \sum_{l=0}^{n-1} {\ti P}_l (\l_2)
{\wti Q}^r_{ln}   \lab{pntits}
\er
where we introduced
\be
{\bar Q}_{nm} \equiv \(H {\wti Q} H^{-1}\)_{nm}
 \quad \mbox{with} \quad H_{nm} = h_n \d_{nm}
\lab{defqbar}
\ee
Define now a wave function
\be
\Psi_n \( t,{\ti t}, \l\) = P_n (\l_1) e^{V(t,\l_1)}
\lab{psipnv}
\ee
{}From definition of $\Psi_n$ and \rf{recurpo}, \rf{pntr} and \rf{pntits}
we obtain the following eigenvalue problem:
\br
\l_1 \Psi &=& Q \Psi \qquad ;\qquad   \lab{lpsiq}\\
\partder{\Psi}{t_r} &=& Q^r_{(+)} \Psi \lab{tpsiq}\\
\partder{\Psi}{{\ti t}_s} &=& - {\bar Q}^s_{-} \Psi \lab{titpsiq}
\er
where $\Psi$ is a semi-infinite column
$\( \ldots, \Psi_n, \Psi_{n+1}, \ldots \)^{T}$ and
$Q$ and ${\bar Q}$ are semi-infinite matrices, {\sl i.e.}, with indices
running from $0$ to $\infty$,
Furthermore we adhere to the following notation:
the subscripts $-/+$ denote lower/upper triangular parts of the matrix,
whereas $(+)/(-)$ denote upper/lower triangular plus diagonal parts.

In addition we also have a relation:
\be
\partder{\Psi}{\l_1}= -g {\bar Q} \Psi \lab{lpsibarq}
\ee
which can be derived using identity:
\be
0 =\int d \l_1 d \l_2 \partder{}{\l_i} \exp \( V (\l_1) + {\ti V} (\l_2) +
 g \l_1 \l_2 \) P_n (\l_1) {\ti P}_m (\l_2) \qquad i =1,2
\lab{dortho}
\ee
\subsection{Two-Matrix Model as a Discrete Linear System}
The compatibility of the eigenvalue problem \rf{lpsiq}-\rf{lpsibarq}
\ct{BX9212,enjoy} gives rise to a discrete linear system which we shall call
the {\em constrained} generalized Toda lattice hierarchy:
\br
\partder{}{t_r} Q = \llb Q^r_{(+)} , Q \rrb  \quad , \quad
\partder{}{t_r} {\bar Q} = \llb Q^r_{(+)} , {\bar Q} \rrb \quad , \quad
r=1,\ldots , p_1  \lab{L-3} \\
\partder{}{{\ti t}_s} Q = \llb Q , {\bar Q}^s_{-} \rrb     \quad , \quad
\partder{}{{\ti t}_s} {\bar Q} = \llb {\bar Q} , {\bar Q}^s_{-} \rrb
\quad ,\quad s=1,\ldots ,p_2    \lab{L-4}\\
- g \llb Q , {\bar Q} \rrb = \one   \phanta   \lab{string-eq}
\er
As will be shown later the lattice system \rf{Tod-1}--\rf{Tod-3} can be
identified with the ${\ti t}_1 \equiv x$ evolution equations of
the above discrete linear system and therefore provides a direct link
between matrix models and the integrable hierarchies \ct{integr-matrix}.

In what follows it will be convenient to define an explicit
parametrization of matrices $Q$ and ${\bar Q}$.
We choose the following parametrization which is
consistent with \rf{qtiqone}:
\br
Q_{nn} = a_0 (n) \quad , \quad Q_{n,n+1} =1 \quad ,\quad
Q_{n,n-k} = a_k (n) \quad k=1,\ldots , p_2 -1   \nonu  \\
Q_{nm} = 0 \quad {\rm for} \;\;\; m-n \geq 2 \;\; ,\;\; n-m \geq p_2 \phanta
\lab{param-1}  \\
{\bar Q}_{nn} = b_0 (n) \quad , \quad {\bar Q}_{n,n-1} = R_n \quad , \quad
{\bar Q}_{n,n+k} = b_k (n) R_{n+1}^{-1} \cdots R_{n+k}^{-1}
\quad k=1,\ldots ,p_1 -1    \nonu  \\
{\bar Q}_{nm} = 0 \quad {\rm for} \;\;\; n-m \geq 2 \;\; ,\;\; m-n \geq p_1
\phanta    \lab{param-2}
\er
In most examples used in this section we work
with the number $p_2 =3$ in \rf{param-1}, whereas the
number $p_1$ in \rf{param-2} remains finite or $\infty$ \foot{Both
numbers $p_{1,2}$ indicating the number of non-zero diagonals, outside the
main one, of the matrices ${\bar Q}$ and $Q$ are related with the polynomial
orders of the corresponding string two-matrix model potentials, whereas the
constant $g$ in \rf{string-eq} denotes the coupling parameter between the
two random matrices.}.

Let us consider for the moment the first evolution parameters $t_1 ,{\ti t}_1$
as coordinates of a two-dimensional space, {\sl i.e.}, ${\ti t}_1 \equiv
{\ti x}$ and $t_1 \equiv y$, so all modes $a_k (n) , b_k (n)$ and $ R_n $
depend on $\( {\ti x},y ;t_2 ,\ldots ,t_{p_1};{\ti t}_2 ,\ldots
,{\ti t}_{p_2} \)$ .

The second lattice equation of motion \rf{L-4} for $s=1$, using
parametrization \rf{param-2}, gives :
\br
\pa_{\ti x} R_n &=& \! R_n \( b_0 (n) - b_0 (n-1) \)\quad,\quad
\pa_{\ti x} b_0 (n)= b_1 (n) - b_1 (n-1)  \phantom{aa}  \lab{motionx-0} \\
\pa_{\ti x} \( \frac{b_k (n)}{R_{n+1} \ldots R_{n+k}} \)& =&
\frac{b_{k+1}(n) - b_{k+1}(n-1)}{R_{n+1} \ldots R_{n+k}}  \quad , \;\;
k \geq 2   \lab{motionx-k}
\er
Similarly, the second lattice equation of motion \rf{L-3} for $r=1$ gives :
\be
\pa_y b_0 (n) = R_{n+1} - R_n    \quad,\quad
\pa_y b_k (n) = R_{n+1} b_{k-1}(n+1) - R_{n+k} b_{k-1} (n)
\lab{motiony-k}
\ee
for $k \geq 1$.
{}From the above equations one can express all $b_k (n \pm \ell) \;\; , \;
k \geq 2\,$ and $ R_{n \pm \ell}\,$ ($\ell$ -- arbitrary integer) as
functionals of $b_0 (n) , b_1 (n)$ at a {\em fixed} lattice site $n$.

Furthermore, the lattice equations of motion \rf{L-3}-\rf{L-4} for
$r=1, s=1$ read explicitly:
\be
\pa_{\ti x} a_0 (n) = R_{n+1} - R_n \quad,\quad
\pa_{\ti x} a_k (n) = R_{n-k+1} a_{k-1}(n) - R_n a_{k-1}(n-1) \lab{motionxx-k}
\ee
(with $k \geq 1$) and
\be
\pa_y a_0 (n) = a_1 (n+1) - a_1 (n) \quad,\quad
\pa_y \( \frac{a_k (n)}{R_n \ldots R_{n-k+1}}\) =
\frac{a_{k+1}(n+1) - a_{k+1}(n)}{R_n \ldots R_{n-k+1}}
\lab{motionyy-k}
\ee
with $k \geq 1 $.
Following \rf{motionx-0}, \rf{motiony-k}, \rf{motionxx-k} we obtain the
``duality'' relations:
\be
\pa_y b_1 (n) = \pa_{\ti x} R_{n+1}
\quad , \quad\pa_{\ti x} a_0 (n) = \pa_y b_0 (n)
\quad , \quad \pa_{\ti x} a_1 (n) = \pa_y R_n
\lab{dual}
\ee
{}From the above one gets the two-dimensional Toda lattice equation:
\be
\pa_y \ln R_n = a_0 (n) - a_0 (n-1)
\quad \to \quad
\pa_{\ti x} \pa_y \ln R_n = R_{n+1}  - 2 R_{n} + R_{n-1}  \lab{motion-00}
\ee

Eqs.\rf{motionxx-k}--\rf{motion-00} allow to express all $a_k (n \pm \ell )
\;\; ,\; k \geq 1$ and $R_{n \pm \ell}$ as functionals of $a_0 (n)$ and
$R_n$ (or $a_1 (n)$ instead) at a fixed lattice site $n$. Furthermore,
due to eqs.\rf{motionxx-k} and \rf{motiony-k} for $k=1$, all
matrix elements of $Q$ and ${\bar Q}$ are functionals of $b_0 (n) , b_1 (n)$
at a fixed lattice site $n$.
Alternatively, due to \rf{dual} we can consider $a_0 (n)$ and $R_{n+1}$ as
independent functions instead of $b_0 (n),b_1 (n)$.

Let us also add the explicit expressions for the flow
eqs. for $b_0 (n), b_1 (n), R_{n+1}$ resulting from \rf{L-3} and \rf{L-4} :
\br
\partder{}{{\ti t}_s}b_0 (n) \eq \pa_{\ti x} \( {\bar Q}^s\)_{nn} \; ,\;
\partder{}{{\ti t}_s}b_1 (n) =
\pa_{\ti x} \Bigl\lb R_{n+1}\( {\bar Q}^s\)_{n,n+1}\Bigr\rb    \;,\;
\partder{}{{\ti t}_s}R_{n+1} = \pa_{\ti x} \( {\bar Q}^s\)_{n+1,n}
\phantom{aa}\lab{t-s-eqs}\\
\partder{}{t_r}b_0 (n) \eq \pa_{\ti x} \( Q^r\)_{nn}  \; ,\;
\partder{}{t_r}b_1 (n) =
\pa_{\ti x} \Bigl\lb R_{n+1}\( Q^r\)_{n,n+1}\Bigr\rb  \; ,\;
\partder{}{t_r}R_{n+1} = \pa_{\ti x} \( Q^r\)_{n+1,n} \phantom{aa}
\lab{t-r-eqs}
\er
\subsection{String Equation}
Note the presence of the non-evolution constraint equation \rf{string-eq},
which is called a string equation.
The lattice equations for the matrix elements $a_k (n)$ of $Q$
(the first eqs.\rf{L-3} and \rf{L-4}) can be solved explicitly as functionals
of the matrix elements of ${\bar Q}$ :
\be
Q_{(-)} = \sum_{s=0}^{p_2 -1} \a_s {\bar Q}^s_{(-)}  \qquad ; \quad
\a_s \equiv - (s+1) \frac{{\ti t}_{s+1}}{g}           \lab{2-3}
\ee
Equation \rf{2-3} follows simply from \rf{dortho} with $i=2$.

Note also that there is a complete duality
under $\, p_1 \longleftrightarrow p_2\,$ when the order $p_1$ of the first
matrix potential in \rf{2-1} is also finite.
For instance due to this duality we can obtain the analog of eq. \rf{2-3}:
\be
{\bar Q}_{(+)} = \sum_{r=0}^{p_1-1} \b_r Q^r_{(+)}  \qquad ; \quad
\b_r \equiv - (r+1) \frac{t_{r+1}}{g}           \lab{2-3ana}
\ee
by interchanging
$\, p_1 \longleftrightarrow p_2 \; ,\; {\ti t}_s \longleftrightarrow t_r \; ,\;
Q_{(-)} \longleftrightarrow {\bar Q}_{(+)}$.
Equation \rf{2-3ana} follows simply from \rf{dortho} with $i=1$.

The first equations of \rf{motionxx-k} and \rf{motion-00},
imply the following two additional constraints on the independent functions
$b_0 (n)$ and $R_{n+1}$ (or $b_1 (n)$)
\be
\pa_y b_0 (n) = \pa_x \( \sum_{s=0}^{p_2 -1}\a_s {\bar Q}^s_{nn}\) \quad
;\quad
\pa_y R_{n+1} = \pa_x \( \sum_{s=1}^{p_2 -1}\a_s {\bar Q}^s_{n+1,n}\)
\lab{constr-0}
\ee
In fact, eqs. \rf{constr-0} are nothing but the component form
of the string equation \rf{string-eq}.
Indeed upon using \rf{2-3} and \rf{L-3}, \rf{L-4}, eq. \rf{string-eq}
can be rewritten in the following form:
\be
\(\pa_y - \sum_{s=1}^{p_2 -1} \a_s \pa/\pa{\ti t}_s \){\bar Q} =
-{1\over g}\one
\lab{streq}
\ee
or equivalently
\br
&&\(\pa_y - \sum_{s=1}^{p_2 -1}\a_s \pa/\pa{\ti t}_s \) b_0 (n) = - {1\over g}
\nonu \\
&&\(\pa_y - \sum_{s=1}^{p_2 -1}\a_s \pa/\pa{\ti t}_s \) R_{n+1} = 0 \quad ,
\quad
\(\pa_y - \sum_{s=1}^{p_2 -1}\a_s \pa/\pa{\ti t}_s \) b_k (n) = 0
\lab{string-eq+}
\er
for $k \geq 1$. Now, inserting \rf{t-s-eqs} into \rf{constr-0} we find that
the latter two equations precisely coincide with the $nn$ and $n+1,n$
component of the string eq.\rf{string-eq+}.
Hence the string equation would amount to identifying the flow $\pa / \pa y
= \pa / \pa t_1 $ with the flow
\be
\partder{}{{\hat t}_{p_2-1}} \equiv \sum_{s=1}^{p_2 -1}\a_s
\partder{}{{\ti t}_s }
\lab{defhtp2}
\ee
if not for the constant $- 1/ g$ on the right hand side of \rf{string-eq+}.
For this reason it is more convenient to make a change of variables and
correspondingly introduce another matrix ${\hat Q}$:
\be
{\hat Q}^{p_2 -1}_{(-)} \, \equiv \, \sum_{s=0}^{p_2 -1} \a_s {\bar Q}^s_{(-)}
\lab{tatko}
\ee
We now generalize \rf{tatko} to:
\be
{\hat Q}^s_{(-)} = \sum_{\s =0}^s
\g_{s\s} {\bar Q}^{\s}_{(-)} \qquad s=1,\ldots , p_2
\lab{tatko-gen}
\ee
and parametrize ${\hat Q}$ as in \rf{param-2})
with matrix elements ${\hat R}_n, {\hat b}_k (n)$
in place of $R_n, b_k (n)$.
Coefficients $\g_{s\s}$
are simply expressed through $\a_s \equiv \g_{p_2 -1,s}$ ($\g_{00} \equiv 1$)
as:
\br
\g_{ss} \eq \( \g_{11}\)^s \;, \;
\g_{s,s-1} = {s\over 2} \(\g_{11}\)^{s-2} \g_{21}  \;, \;
\g_{s,s-2} = \(\g_{11}\)^{s-4} \llb \frac{s(s-3)}{8} \(\g_{21}\)^2 +
{s \over 3} \g_{11} \g_{31} \rrb    \phantom{aaa} \lab{tatko-2-a}  \\
\g_{11} \! &\equiv&\! \( \a_{p_2 -1} \)^{1\over {p_2 -1}} \quad ,\quad
\g_{21} \equiv {2\over {p_2 -1}} \frac{\a_{p_2 -2}}{\(\a_{p_2 -1}\)^{{p_2 -3}
\over {p_2 -1}}}  \nonu \\
\g_{31} \! &\equiv&\!  {3\over {p_2 -1}} \llb
\frac{\a_{p_2 -3}}{\(\a_{p_2 -1}\)^{{p_2 -4}
\over {p_2 -1}}} - \frac{p_2 -4}{2\( p_2 -1\)}
\frac{\a_{p_2 -2}^2}{\(\a_{p_2 -1}\)^{{2p_2 -5}\over {p_2 -1}}}\rrb
\lab{tatko-2}
\er
and for the ${\hat Q}$-matrix elements we obtain:
\be
{\hat R}_n = \g_{11} R_n \quad , \quad {\hat b}_0 (n) = \g_{11} b_0 (n) +
\frac{\g_{21}}{2\g_{11}} \quad ,\quad {\hat b}_1 (n) = \g_{11}^2 b_1 (n) +
\frac{\g_{31}}{3\g_{11}} - \( \frac{\g_{21}}{2\g_{11}} \)^2  \lab{tatko-1}
\ee
etc..

We can now rewrite the string equation in the ``hatted" variables
from eq. \rf{tatko-1}.
First we note that the explicit calculation based on \rf{tatko-2} gives
\be
\partder{}{{\hat t}_{p_2-1}} \g_{s,s-1} = - { s \o g} \g_{s,s}
\quad \mbox{and} \quad
\partder{}{{\hat t}_{p_2-1}} \g_{s,s-2} = - { s+1 \o 2 g} \g_{s,s-1}
\lab{tgss}
\ee
Especially, we find
\be
\partder{}{{\hat t}_{p_2-1}} \g_{21} = - { 2 \o g} \g_{11}^2
\lab{tg21}
\ee
and
\be
\partder{}{{\hat t}_{p_2-1}} \llb \frac{\g_{31}}{3\g_{11}} -
\( \frac{\g_{21}}{2\g_{11}} \)^2  \rrb =0
\ee
Hence we easily get
\be
\(\partder{}{t_1}- \partder{}{{\hat t}_{p_2-1}} \) {\hat b}_i (n) =0
 \qquad i=1,2
 \lab{bone}
\ee
and the same result for ${\hat R}_n$.
This allows us to postulate a general statement:
\be
\(\partder{}{t_1}- \partder{}{{\hat t}_{p_2-1}} \) {\hat Q}=0
\lab{hatqone}
\ee
Accordingly in the new ``hatted" framework the string equation boils down
to identification of the flow $\pa / \pa t_1 $ with  $\pa / \pa
{\hat t}_{p_2-1}$ flow.

Consider now an arbitrary matrix $X ( Q, {\bar Q})$. From equations
\rf{L-4}, \rf{L-3} and \rf{2-3} we find
\be
\partder{X ( Q, {\bar Q})}{{\hat t}_{p_2-1}} = \sbr{X}{Q_{-}}
= \sbr{Q_{(+)}}{X} + \sbr{X}{Q}=
\partder{X}{t_1} + \sbr{X}{Q}
\lab{xqq}
\ee
Comparing with \rf{hatqone} we find that $\sbr{{\hat Q}}{Q}=0$.
It is therefore plausible to supplement \rf{tatko} by condition
\be
{\hat Q}^{p_2 -1}_{(+)} = I_{+}  \longrightarrow {\hat Q}^{p_2 -1} = Q
\lab{tatko-s}
\ee
with ${\(I_{+}\)}_{nm}= \d_{n+1,m}$. The last equality follows from \rf{2-3}.
Let us also introduce the new subset of evolution parameters
$\lcurl {\hat t}_s \rcurl$ through the equations:
\be
\partder{}{{\hat t}_s} \, \equiv \, \sum_{\s =1}^s \g_{s\s}
\partder{}{{\ti t}_{\s}} \quad \;,\quad \; s=1,\ldots, p_2      \lab{tatko-a}
\ee
which for $s=p_2-1$ reproduces \rf{defhtp2}.
Now, taking into account \rf{tatko}, \rf{tatko-a} and condition
\rf{tatko-s}, all constrained Toda
lattice eqs.\rf{L-3}--\rf{string-eq} can be re-expressed as a
single set of the flow equations for the matrix ${\hat Q}$ :
\be
\partder{}{{\hat t}_s} {\hat Q} = \llb {\hat Q} , {\hat Q}^s_{-} \rrb \quad
\;,\; \;\; s=1,\ldots,p_2 , 2(p_2 -1), 3(p_2 -1), \ldots, p_1 (p_2 -1)
   \lab{L-4-hat}
\ee
with the identification $t_r \equiv {\hat t}_{r(p_2 -1)}$,
The associated linear problem reads:
\be
{\hat Q}_{nm} {\hat \psi}_m = \l {\hat \psi}_n  \quad , \quad
\partder{}{{\hat t}_s} {\hat \psi}_n = - \({\hat Q}^s_{-}\)_{nm} {\hat \psi}_m
\lab{L-hat}
\ee
{}From the fact that ${\hat Q}$ has the same form as ${\bar Q}$ up to the
``hatting"
we can deduce that ${\hat Q}$ satisfies the Zakharov-Shabat equation:
\be
\partder{}{{\hat t}_r} {\hat Q}^s_{-} - \partder{}{{\hat t}_s} {\hat Q}^r_{-}
+ \llb {\hat Q}^r_{-} , {\hat Q}^s_{-} \rrb =0
\lab{hat-zs}
\ee
Recalling associated linear problem \rf{L-hat}
we see that \rf{hat-zs} translates into:
\be
\llb \partder{}{{\hat t}_r} , \partder{}{{\hat t}_{s}} \rrb {\hat \psi}= 0
\lab{commute}
\ee
establishing the commutativity of flows.
In fact an explicit calculation
for the simplest nontrivial case of $p_2=3$ gives indeed \rf{commute}
for $r,s =1,2,3$ !

Recall now the equation \rf{2-3ana}. Together with the string equation
\rf{string-eq} it imposes a following condition on $Q$:
\be
\( \sum_{r=1}^{p_1-1} \b_r \partder{}{t_{r}} - \partder{}{{\ti t}_{1}} \)
Q = \( \sum_{r=1}^{p_1-1} \b_r \partder{}{t_{r}} - \g^{-1}_{11}
\partder{}{{\hat t}_{1}} \) Q = {1 \o g}
\lab{condana}
\ee
Due to identification between ${\hat Q}$ and $Q$ modes in \rf{tatko-s}
this extra condition translates into a condition for ${\hat Q}^{p_2-1}$.

We obtain from \rf{L-4-hat},\rf{L-hat} the following continuum
Lax problem at a fixed lattice site $n$:
\br
\l^{p_2-1} {\hat \psi}_n &=& {\hat L}(n) {\hat \psi}_n \quad ,\quad
\partder{}{{\hat t}_s} {\hat \psi}_n = - {\hat \cL}_s (n) {\hat \psi}_n
\lab{L-cont-h} \\
\partder{}{{\hat t}_s} {\hat L}(n) \eq \sbr{{\hat L}(n)}{{\hat \cL}_s (n)}
\quad , \;\; s=1,\ldots,p_2 , 2(p_2 -1), 3(p_2 -1), \ldots p_1 (p_2 -1)
\lab{L-cont-1-h}  \\
{\hat L}(n) \!&\equiv&\! -D_x^{-1}\, {\hat R}_{n+1} + {\hat Q}^{p_2 -1}_{nn}
+ \sum_{k=1}^{p_2 -1}
\frac{(-1)^k{\hat Q}^{p_2 -1}_{n,n-k}}{{\hat R}_n\ldots {\hat R}_{n-k+1}}
\Bigl( D_x - \pa_x \ln \( {\hat R}_n \ldots {\hat R}_{n-k+2}\)\Bigr)
\times \ldots \nonu\\
&& \ldots \times \Bigl( D_x - \pa_x \ln {\hat R}_n \Bigr)\,D_x \lab{Ln-x} \\
{\hat \cL}_s (n) &\equiv& \sum_{k=1}^s
\frac{(-1)^k {\hat Q}^{s}_{n,n-k}}{{\hat R}_n\ldots {\hat R}_{n-k+1}}
\Bigl( D_x - \pa_x \ln \( {\hat R}_n \ldots {\hat R}_{n-k+2}\)\Bigr)
\ldots D_x  \lab{Ls-x}
\er
where all coefficients are expressed in terms of ${\hat R}_{n+1},
{\hat b}_0 (n),\ldots ,{\hat b}_{p_1 -2}(n)$ at a fixed site $n$ through the
${\hat t}_1 \equiv x$ lattice equations of motion \rf{L-4-hat}. Up to a gauge
transformation and conjugation, the explicit form of ${\hat L}(n)$ reads:
\be
e^{\int {\hat b}_0 (n)} \( {\hat L}(n)\)^{\ast} e^{-\int {\hat b}_0 (n)} =
D_x^{p_2 -1} + \( p_2 -1\) {\hat b}_1 (n) D_x^{p_2 -3} + \cdots +
{\hat R}_{n+1} \( D_x - {\hat b}_0 (n) \)^{-1}    \lab{hat-Ln}
\ee
Eqs.\rf{L-cont-h}--\rf{Ls-x} are the continuum analogs of the constrained Toda
lattice Lax equations \rf{lpsiq}--\rf{lpsibarq} and \rf{L-3}--\rf{string-eq}
without taking any continuum
(double-scaling) limit. Let us particularly stress that they explicitly
incorporate the whole information from the matrix-model string equation
\rf{string-eq} through \rf{2-3} and \rf{tatko}--\rf{tatko-1}
which were used in their derivation.

Consider a case $p_2 =3$. The Lax  operator ${\hat L}(n) $ from \rf{Ln-x}
is given by:
\be
{\hat L}(n) = -D^{-1}\, {\hat R}_{n+1} +  2{\hat b}_1 (n) +
{\hat b}_n^2 (n) - \pa_x {\hat b}_0 (n)
- 2 {\hat b}_0 (n) D + D^2   \lab{lax-bar}
\ee
The corresponding lattice equations then read (we write down explicitly
only the ${\hat t}_1 \equiv x$ and ${\hat t}_2$
evolution equations ) :
\br
\pa_x \ln {\hat R}_{n+1} \eq {\hat b}_0 (n+1) - {\hat b}_0 (n) \quad ,\quad
{\hat b}_1 (n) - {\hat b}_1 (n-1) = \pa_x {\hat b}_0 (n)  \lab{x-motion-1} \\
{\hat R}_{n+1} - {\hat R}_n \eq \pa_x \( {\hat b}_0^2 (n) + {\hat b}_1 (n) +
{\hat b}_1 (n-1) \)    \lab{x-motion-2} \\
\partder{}{{\hat t}_2} {\hat R}_{n+1} \eq\pa_x \Bigl\lb \pa_x {\hat R}_{n+1} +
2{\hat b}_0 (n) {\hat R}_{n+1} \Bigr\rb   \phanta  \lab{1+1-aa} \\
\partder{}{{\hat t}_2} {\hat b}_0 (n) \eq \pa_x \Bigl\lb 2 {\hat b}_1 (n) +
{\hat b}_0^2 (n) - \pa_x {\hat b}_0 (n) \Bigr\rb   \quad ,\quad
\partder{}{{\hat t}_2} {\hat b}_1 (n) =  \pa_x {\hat R}_{n+1}   \lab{1+1-cc}
\er
Now, we observe that the system of Darboux-\Back ~equations for $SL(3,1)$
\cKP hierarchy \rf{Tod-1}--\rf{Tod-3} exactly coincides upon identification:
\be
B^{(n)}= {\hat b}_0 (n-1) \quad ,\quad u^{(n)} = 2 {\hat b}_1 (n-1) \quad
,\quad
A^{(n)}= {\hat R}_n  \lab{Tod-5}
\ee
with the $x\equiv {\hat t}_1 $ constrained Toda lattice evolution equations
\rf{x-motion-1}--\rf{x-motion-2}.
Also, the higher Toda lattice evolution parameters
can be identified with the following subset of evolution parameters of the
$SL(p_2 ,1)$ KP-KdV hierarchy \rf{sol-9} \ct{enjoy,office} :
\be
{\hat t}_s \simeq t_s^{KP-KdV} \quad s=2,\ldots ,p_2 \quad ; \quad
t_r \simeq t_{r(p_2 -1)}^{KP-KdV} \quad r=1,\ldots ,p_1   \lab{param}
\ee
the second identification resulting from \rf{tatko}.

In particular, excluding ${\hat b}_0 (n) \equiv B^{(n+1)}$ and
${\hat b}_1 (n) \equiv \h u^{(n+1)}$ in \rf{Tod-3} using
\rf{x-motion-1}--\rf{1+1-cc}, we obtain the two-dimensional Toda lattice
equation for $A^{(n)}\equiv {\hat R}_n$ :
\be
\pa_x \pa_{{\hat t}_2} \ln A^{(n)} = A^{(n+1)}  - 2 A^{(n)} + A^{(n-1)}
\lab{motion-toda}
\ee
\subsection{Partition Function of the Two-Matrix String Model}
The partition function $Z_N$ of the two-matrix string model is simply
expressed in terms of the ${\bar Q}$ matrix element $b_1 (N-1)$ at
the Toda lattice site $N-1$, where $N$ indicates the size ($N \times N$) of
the pertinent random matrices: $\pa_x^2 \ln Z_N = b_1 (N-1)$
(cf. \ct{BX9212,enjoy}). Thus, using \rf{tauok} and \rf{sol-6} together with
\rf{param}, and accounting for the relations \rf{tatko-1}--\rf{tatko-2},
 we obtain the following exact
solution at {\em finite} $N$ for the two-matrix model partition function:
\br
Z_N &= & W_N \llb {\hat \Phi}^{(0)}, \pa_{{\hat t}_{p_2 -1}} {\hat \Phi}^{(0)},
\ldots , \pa_{{\hat t}_{p_2 -1}}^{N-1} {\hat \Phi}^{(0)} \rrb\,
\exp - N \int^{{\hat t}_1}   \( \frac{\g_{21}}{2\g_{11}}\)
\lab{sol-3-matr}  \\
{\hat \Phi}^{(0)} &=& \int_{\Gamma} \frac{d\l}{2\pi} c(\l )
e^{{\hat \xi} \( \l ,\{ {\hat t},t\}\)} \quad ,\;\;
{\hat \xi}\( \l ,\{ {\hat t},t\}\) \equiv \sum_{s=1}^{p_2} \l^s {\hat t}_s
+ \sum_{r=2}^{p_1} \l^{r(p_2 -1)} t_r     \lab{sol-6-matr}
\er
where $ W_N \llb \ldots \rrb \equiv \det \(  \pa_x^{i-1}
\pa_{{\hat t}_{p_2-1}}^{j-1} {\hat \Phi}^{(0)} \)_{1 \leq i,j \leq N}$
with $x \equiv {\hat t}_1$. The $\g$-coefficients are defined in
\rf{tatko-2}. The ``density'' function $c(\l )$ in \rf{sol-6-matr} is
determined from matching the expression for ${\hat \Phi}^{(0)}$:
$\pa_x \ln {\hat \Phi}^{(0)} = {\hat b}_0 (0) = \g_{11} b_0 (0) +
\g_{21}/2\g_{11}$ (cf. \rf{Tod-4},\rf{Tod-5},\rf{tatko-1}),
with the expression for $b_0 (0)$ in the orthogonal-polynomial
formalism \ct{BX9212}:
\br
&&\int_{\Gamma} \frac{d\l}{2\pi} c(\l ) \exp \( \sum_{s=1}^{p_2}
\l^s {\hat t}_s + \sum_{r=2}^{p_1} \l^{r(p_2 -1)} t_r \)  \lab{tatko-3}\\
\eq \exp \( \int^{{\ti t}_1} \frac{\g_{21}}{2 \g_{11}^2}\)
 \int_\Gamma \int_\Gamma d\l_1 d\l_2 \,
\exp \(\sum_{r=1}^{p_1} \l_1^r t_r + \sum_{s=1}^{p_2} \l_2^s {\ti t}_s +
g \l_1 \l_2 \)\Bgv_{t_1 = {\hat t}_{p_2 -1} ({\ti t})}  \nonu
\er
One can easily check that $W_N$ from \rf{sol-3-matr} satisfies the
two-dimensional Toda lattice equation $\pa^2 \ln W_N / \pa {\hat t}_1
\pa {\hat t}_{p_2 - 1} = W_{N+1}W_{N-1} / W_{N}^2$.
Furthermore $Z_N \g_{11}^{-N(N-1)/2}$ provides a solution to the
two-dimensional Toda lattice equation
$\pa^2 \ln Z_N / \pa t_1
\pa {\ti t}_{1} = Z_{N+1}Z_{N-1} / Z_{N}^2$ underlying the two-matrix
model.
This shows relevance of methods based on the Darboux-B\"{a}cklund
transformations for obtaining partition functions of the multi-matrix models.

\newpage
\appendix
\sect{Schur Polynomials}
\name
The Schur polynomials $p (x_1,x_2, \ldots)$ are polynomials of the
multi-variable $ x \equiv (x_1,x_2, \ldots)$ defined by the generating
function:
\be
\exp \( \sumi{k=1} x_k z^k\) = \sumi{k=0} p_k (x) z^k  \lab{schurdef}
\ee
and $p_k (x) =0$ for $k <1$.

In particular
\be
p_0 (x) =1\;, \; p_1  (x) =x_1\; , \; p_2 (x) =x_1^2/2 + x_2
\; , \; p_3 (x) =x_1^3/6 + x_1 x_2 + x_3
\lab{schurex}
\ee
and generally $ p_n (x) =x_1^n/n!  + \ldots +  x_n$. Also
$(\pa / \pa x_m)\, p_n (x) = p_{n-m} (x) $.

We need the following recurrence relation.

\lemma
\be
m p_m \( {\ti x} \) = \sum_{k=1}^m x_k p_{m-k} \( {\ti x}\)
\lab{schur-le}
\ee
where ${\ti x }\equiv \( x_1, {x_2 \o 2}, {x_3 \o 3},  \ldots \)$

\proof
{}From definition \rf{schurdef} we get
\be
\exp \( \sumi{k=1} {x_k \o k} z^k\) = \sumi{k=0} p_k ({\ti x}) z^k
\lab{schurpra}
\ee
Multiplying on both sides by $\sumi{l=0} x_l z_l$ we obtain
\br
\sumi{m=0} \( \sum_{k=1}^m  x_k p_{m-k} ({\ti x })\) z^m
&=& \sumi{l=0} x_l z_l \exp \( \sumi{k=1} {x_k \o k} z^k\)
= z {d \o dz} \exp \( \sumi{k=1} {x_k \o k} z^k\) \nonu\\
&=& z {d \o dz}  \, \sumi{m=0} p_m ({\ti x }) z^m
= \sumi{m=0} m p_m ({\ti x }) z^m  \lab{schurprb}
\er
the final result follows now by comparing coefficients of $z^m$.

Especially taking $x_k = - \pa /\pa t_k$ in \rf{schur-le}
we get
\be
m p_m \( - {\wti \pa} \) = - \sum_{k=1}^m
\partder{p_{m-k} \(- {\ti \pa}\)}{t_k}
= - \sum_{k=1}^{m-1} \partder{p_{m-k} \(- {\ti \pa}\)}{t_k}
-  \partder{}{t_m}
\lab{schur-lea}
\ee
which reproduces \rf{conserv-a} in the text when applied on $\ln \t$ after
an extra $x$-derivative is taken on both sides.
\sect{Wronskian Preliminaries}
\sskp
We list here three basic properties of the Wronskian determinants.
\mskp
{\bf 1)} The derivative $\cd^{\pr}$ of a determinant $\cd$
of order $n$, whose entries are differentiable functions, can be written as:
\be
\cd^{\pr} = \cd_{(1)} + \cd_{(2)} + \ldots + \cd_{(n)}
\lab{derdet}
\ee
where $ \cd_{(i)} $ is obtained from $D$ by differentiating the entries in the
$i$-th row.
\mskp
{\bf 2)} {\sl Jacobi expansion theorem}:
\be
W_{k} \(f\) W_{k-1 }= W_{k } W_{k-1 }^{\pr} \(f\)- W_{k }^{\pr} W_{k-1 } \(f\)
\lab{jac}
\ee
or
\be
 \pa \( { W_{k-1 } (f) \over W_{k}} \) = {W_{k} \(f\) W_{k-1 }
\over W_{k}^2}
\lab{jac-pr}
\ee
where the Wronskians are
\be
W_k \equiv W_k \lb \psi_1, \ldots ,\psi_k \rb=
\det \( {\pa^{i-1} \psi_j \o \pa x^{i-1}} \)_{1 \leq i,j \leq  k}
\ee
and
\be
W_{k-1} \(f \)\equiv W_{k} \lb \psi_1, \ldots ,\psi_{k-1}, f\rb.
\ee
The proof (see also \ct{AvM,am}) of \rf{jac} uses the fact that the left hand
side of \rf{jac-pr} can be written as a linear combination
$\sum_{p=1}^k a_p f^{(p)}$.
Expression for the coefficients $a_p$ (up to multiplication with a common
function) can be obtained by e.g. the Cramer's
rule due to the fact that $\sum_{p=1}^k a_p \psi_i^{(p)}=0 $ for $i=1, \ldots,
k$ as can be verified directly from \rf{jac-pr}.
It is now easy to see that the left hand side must be proportional to
$W_{k} \(f\)$.
The proof follows now by establishing that the terms with $f^{(k-1)}$ on both
sides agree.

Take now a special class of Wronskians
\be
W_n \equiv W_n \lb \psi, \psi^{\pr},
\ldots ,\pa^{n-1} \psi \rb =
\det \({ \pa^{i+j-2} \psi \o \pa x^{i+j-2}}\)_{1 \leq i,j \leq  n} .
\ee
Hence, from \rf{jac} we get :
\be
W_n W_n^{\pr \pr} - W_n^{\pr} W_n^{\pr} =
W_n \, W_{n-1}^{\pr}\(\pa^{n} \psi \) - W_n^{\pr}\, W_{n-1} \(\pa^{n} \psi \)
= W_{n-1} W_{n+1}
\lab{jac-a}
\ee
which can be rewritten as
\be
\pa^2 \ln W_n = {W_{n+1} W_{n-1 } \over W_{n}^2}
\lab{jac-b}
\ee
\mskp
{\bf 3)} Iterative composition of Wronskians:
\be
T_k \, T_{k-1}\, \cdots\, T_1 (f ) \; =\; { W_{k} (f) \over W_k}
\lab{iw}
\ee
where
\be
T_j = { W_{j} \over W_{j-1} } \pa { W_{j-1} \over W_{j} } =
\( \pa + \( \ln { W_{j-1} \over W_{j} } \)^{\pr} \) \quad;\quad W_{0}=1
\lab{transf}
\ee
The proof of \rf{iw} follows by simple iteration of \rf{jac-pr}
(see also the standard references on this subject
\ct{crum,ince,AvM}). For future use let us rewrite \rf{iw} as:
\be
\( \pa + v_k \) \( \pa + v_{k-1} \) \cdots \( \pa + v_1 \) \, f =
{ W_{k} (f) \over W_k} \quad;\qquad v_j \equiv \pa_x \ln { W_{j-1} \over W_j}
\lab{kw}
\ee
or
\be
\( D + v_k \) \( D + v_{k-1} \) \cdots \( D + v_1 \) =
{1 \o W_k} \left\v \begin{array}{cccc}
\psi_1 & \psi_2 & \cdots & 1 \\ \psi_1^{\pr} & \psi_2^{\pr} & \cdots & D\\
\vdots & \vdots & \cdots & \vdots \\
\psi_1^{(k)} & \psi_2^{(k)}& \cdots & D^{k}
\end{array} \right\v
\lab{kw-a}
\ee

\end{document}